\documentclass[12pt,titlepage,letterpaper]{utarticle}

\usepackage{amsmath}
\usepackage{amsfonts}
\usepackage{amssymb}
\usepackage{amsthm}
\usepackage{mathtools}
\usepackage{graphicx}
\usepackage{color}
\usepackage{xparse}
\usepackage[utf8x]{inputenc}
\usepackage{tocloft}
\usepackage{cite}
\usepackage[titletoc,title]{appendix}
\usepackage{hyperref}
\usepackage{longtable}

\definecolor{aqua}{rgb}{0, 1.0, 1.0}
\definecolor{fuschia}{rgb}{1.0, 0, 1.0}
\definecolor{gray}{rgb}{0.502, 0.502, 0.502}
\definecolor{lime}{rgb}{0, 1.0, 0}
\definecolor{maroon}{rgb}{0.502, 0, 0}
\definecolor{navy}{rgb}{0, 0, 0.502}
\definecolor{olive}{rgb}{0.502, 0.502, 0}
\definecolor{purple}{rgb}{0.502, 0, 0.502}
\definecolor{silver}{rgb}{0.753, 0.753, 0.753}
\definecolor{teal}{rgb}{0, 0.502, 0.502}

\makeatletter
\newdimen\itex@wd%
\newdimen\itex@dp%
\newdimen\itex@thd%
\def\itexspace#1#2#3{\itex@wd=#3em%
\itex@wd=0.1\itex@wd%
\itex@dp=#2ex%
\itex@dp=0.1\itex@dp%
\itex@thd=#1ex%
\itex@thd=0.1\itex@thd%
\advance\itex@thd\the\itex@dp%
\makebox[\the\itex@wd]{\rule[-\the\itex@dp]{0cm}{\the\itex@thd}}}
\makeatother

\makeatletter
\newif\if@sup
\newtoks\@sups
\def\append@sup#1{\edef\act{\noexpand\@sups={\the\@sups #1}}\act}%
\def\reset@sup{\@supfalse\@sups={}}%
\def\mk@scripts#1#2{\if #2/ \if@sup ^{\the\@sups}\fi \else%
  \ifx #1_ \if@sup ^{\the\@sups}\reset@sup \fi {}_{#2}%
  \else \append@sup#2 \@suptrue \fi%
  \expandafter\mk@scripts\fi}
\def\tensor#1#2{\reset@sup#1\mk@scripts#2_/}
\def\multiscripts#1#2#3{\reset@sup{}\mk@scripts#1_/#2%
  \reset@sup\mk@scripts#3_/}
\makeatother

\makeatletter
\newbox\slashbox \setbox\slashbox=\hbox{$/$}
\def\itex@pslash#1{\setbox\@tempboxa=\hbox{$#1$}
  \@tempdima=0.5\wd\slashbox \advance\@tempdima 0.5\wd\@tempboxa
  \copy\slashbox \kern-\@tempdima \box\@tempboxa}
\def\slash{\protect\itex@pslash}
\makeatother

\def\clap#1{\hbox to 0pt{\hss#1\hss}}

\let\oldroot\root
\def\root#1#2{\oldroot #1 \of{#2}}
\renewcommand{\sqrt}[2][]{\oldroot #1 \of{#2}}

\DeclareSymbolFont{symbolsC}{U}{txsyc}{m}{n}
\SetSymbolFont{symbolsC}{bold}{U}{txsyc}{bx}{n}
\DeclareFontSubstitution{U}{txsyc}{m}{n}

\DeclareSymbolFont{stmry}{U}{stmry}{m}{n}
\SetSymbolFont{stmry}{bold}{U}{stmry}{b}{n}

\DeclareFontFamily{OMX}{MnSymbolE}{}
\DeclareSymbolFont{mnomx}{OMX}{MnSymbolE}{m}{n}
\SetSymbolFont{mnomx}{bold}{OMX}{MnSymbolE}{b}{n}
\DeclareFontShape{OMX}{MnSymbolE}{m}{n}{
    <-6>  MnSymbolE5
   <6-7>  MnSymbolE6
   <7-8>  MnSymbolE7
   <8-9>  MnSymbolE8
   <9-10> MnSymbolE9
  <10-12> MnSymbolE10
  <12->   MnSymbolE12}{}

\makeatletter
\def\re@DeclareMathSymbol#1#2#3#4{%
    \let#1=\undefined
    \DeclareMathSymbol{#1}{#2}{#3}{#4}}
\re@DeclareMathSymbol{\neArrow}{\mathrel}{symbolsC}{116}
\re@DeclareMathSymbol{\neArr}{\mathrel}{symbolsC}{116}
\re@DeclareMathSymbol{\seArrow}{\mathrel}{symbolsC}{117}
\re@DeclareMathSymbol{\seArr}{\mathrel}{symbolsC}{117}
\re@DeclareMathSymbol{\nwArrow}{\mathrel}{symbolsC}{118}
\re@DeclareMathSymbol{\nwArr}{\mathrel}{symbolsC}{118}
\re@DeclareMathSymbol{\swArrow}{\mathrel}{symbolsC}{119}
\re@DeclareMathSymbol{\swArr}{\mathrel}{symbolsC}{119}
\re@DeclareMathSymbol{\nequiv}{\mathrel}{symbolsC}{46}
\re@DeclareMathSymbol{\Perp}{\mathrel}{symbolsC}{121}
\re@DeclareMathSymbol{\Vbar}{\mathrel}{symbolsC}{121}
\re@DeclareMathSymbol{\sslash}{\mathrel}{stmry}{12}
\re@DeclareMathSymbol{\bigsqcap}{\mathop}{stmry}{"64}
\re@DeclareMathSymbol{\biginterleave}{\mathop}{stmry}{"6}
\re@DeclareMathSymbol{\invamp}{\mathrel}{symbolsC}{77}
\re@DeclareMathSymbol{\parr}{\mathrel}{symbolsC}{77}
\makeatother

\makeatletter
\def\Decl@Mn@Delim#1#2#3#4{%
  \if\relax\noexpand#1%
    \let#1\undefined
  \fi
  \DeclareMathDelimiter{#1}{#2}{#3}{#4}{#3}{#4}}
\def\Decl@Mn@Open#1#2#3{\Decl@Mn@Delim{#1}{\mathopen}{#2}{#3}}
\def\Decl@Mn@Close#1#2#3{\Decl@Mn@Delim{#1}{\mathclose}{#2}{#3}}
\Decl@Mn@Open{\llangle}{mnomx}{'164}
\Decl@Mn@Close{\rrangle}{mnomx}{'171}
\Decl@Mn@Open{\lmoustache}{mnomx}{'245}
\Decl@Mn@Close{\rmoustache}{mnomx}{'244}
\makeatother

\makeatletter
\DeclareRobustCommand\widecheck[1]{{\mathpalette\@widecheck{#1}}}
\def\@widecheck#1#2{%
    \setbox\z@\hbox{\m@th$#1#2$}%
    \setbox\tw@\hbox{\m@th$#1%
       \widehat{%
          \vrule\@width\z@\@height\ht\z@
          \vrule\@height\z@\@width\wd\z@}$}%
    \dp\tw@-\ht\z@
    \@tempdima\ht\z@ \advance\@tempdima2\ht\tw@ \divide\@tempdima\thr@@
    \setbox\tw@\hbox{%
       \raise\@tempdima\hbox{\scalebox{1}[-1]{\lower\@tempdima\box
\tw@}}}%
    {\ooalign{\box\tw@ \cr \box\z@}}}
\makeatother

\makeatletter
\NewDocumentCommand\mathraisebox{moom}{%
\IfNoValueTF{#2}{\def\@temp##1##2{\raisebox{#1}{$\m@th##1##2$}}}{%
\IfNoValueTF{#3}{\def\@temp##1##2{\raisebox{#1}[#2]{$\m@th##1##2$}}%
}{\def\@temp##1##2{\raisebox{#1}[#2][#3]{$\m@th##1##2$}}}}%
\mathpalette\@temp{#4}}
\makeatletter

\makeatletter
\def\udots{\mathinner{\mkern2mu\raise\p@\hbox{.}
\mkern2mu\raise4\p@\hbox{.}\mkern1mu
\raise7\p@\vbox{\kern7\p@\hbox{.}}\mkern1mu}}
\makeatother

\newcommand{\gt}{>}

\theoremstyle{plain}

\theoremstyle{definition}

\theoremstyle{remark}

\numberwithin{equation}{section}

\allowdisplaybreaks



\begin{document}


\preprint{
UTTG--30--15\\
ICTP--SAIFR/2016--XXX\\
}

\title{Tinkertoys for the $\mathbb{Z}_3$-twisted $D_4$ Theory}

\author{Oscar Chacaltana
    \address{
    ICTP South American Institute for\\ Fundamental Research,\\
    Instituto de F\'isica Te\'orica,\\Universidade Estadual Paulista,\\
    01140-070 S\~{a}o Paulo, SP, Brazil\\
    {~}\\
    \email{chacaltana@ift.unesp.br}\\
    },
    Jacques Distler
     \address{
      Theory Group and\\
      Texas Cosmology Center\\
      Department of Physics,\\
      University of Texas at Austin,\\
      Austin, TX 78712, USA \\
      {~}\\
      \email{distler@golem.ph.utexas.edu}\\
      }
      and Anderson Trimm
      \address{
      School of Physics and Astronomy\\
      Center for Theoretical Physics\\
      Seoul National University\\
      Seoul 08826 KOREA\\
      {~}\\
      Fields, Gravity \& Strings\\
      Center for Theoretical Physics\\
      of the Universe\\
      Institute for Basic Sciences\\
      Daejeon 34047 KOREA\\
      {~}\\
      \email{atrimm@physics.utexas.edu}
      }
}
\date{January 10, 2016}

\Abstract{
Among the simple Lie algebras, $D_4$ is distinguished as the unique one whose group of outer-automorphisms is bigger than $\mathbb{Z}_2$. We study the compactifications of the $D_4$ (2,0) Theory on a punctured Riemann surface, $C$, with outer-automorphism twists around cycles of $C$ lying in $\mathbb{Z}_3\subset \text{Aut}(D_4)= S_3$. The resulting 4D $\mathcal{N}=2$ SCFTs have a number of new and interesting properties. As byproduct, we discover a \emph{new} rank-1 $\mathcal{N}=2$ SCFT with flavour symmetry group $SU(4)$.
}

\maketitle

\tocloftpagestyle{empty}
\tableofcontents
\vfill
\newpage
\setcounter{page}{1}

\section{Introduction}\label{introduction}

In recent years, remarkable progress has been made in the study of $4D$ $\mathcal{N}=2$ superconformal field theories by realizing them as partially-twisted compactifications of $6D$ $(2,0)$ theories of type $\mathfrak{j}=A,D,E$ on a punctured Riemann surface, $C$ \cite{Gaiotto:2009hg,Gaiotto:2009we,Alday:2009aq,Gaiotto:2009gz,Gadde:2010te,Gadde:2011ik,Beem:2014rza}. In addition to ordinary $\mathcal{N}=2$ gauge theories, this class of theories (sometimes called ``class $\mathcal{S}$'') contains many strongly-interacting SCFTs, with no known Lagrangian description. An even larger class of theories can be constructed by allowing twists, along nontrivial cycles of $C$, by the action of the outer-automorphism group of $\mathfrak{j}$ \cite{Tachikawa:2010vg,Chacaltana:2012zy}. Doing so introduces a new class of (``twisted'') punctures, labeled by a nilpotent orbit in $\mathfrak{g}$, the Langlands-dual of the invariant subalgebra, $\mathfrak{g}^\vee\subset\mathfrak{j}$.  It also introduces new (``twisted'') cylinders, with gauge groups, $H\subset G$. In \cite{Chacaltana:2012ch,Chacaltana:2013oka,Chacaltana:2014ica,Chacaltana:2015bna,Chacaltana:2014nya}, we studied the $\mathbb{Z}_2$-twisted versions of these theories.

Among the surprising features of the twisted case is that not all boundaries of the moduli space of punctured curves correspond to weakly-coupled gauge theories. Rather, the gauge theory moduli space is, in general, a branched cover of $\overline{M}_{g,n}$, branched over the boundary, and certain components of the boundary of ${M}_{g,n}$ correspond to strongly-coupled gauge theories ({\it i.e.}, interior points in the gauge theory moduli space).

In the present work we turn to the $\mathbb{Z}_3$-twisted $D_4$ theory. $D_4$ is the unique case where the group of outer automorphisms is bigger than $\mathbb{Z}_2$: the full outer-automorphism group is $S_3$. We studied the $\mathbb{Z}_2$-twisted theory in \cite{Chacaltana:2013oka}. For reasons explained in \S\ref{intro}, studying the full nonabelian group of twists is too ambitious for the present work, so we content ourselves with the other abelian subgroup, $\mathbb{Z}_3\subset S_3$.

Perhaps the most interesting byproduct of this work is the construction of a new isolated rank-1 $\mathcal{N}=2$ SCFT, in \S\ref{newSCFT}. The theory has global symmetry ${SU(4)}_{14}$, and a 1-dimensional Coulomb branch parameterized by $u$, where $\Delta(u)=6$.

\section{{Twists: Abelian and non-Abelian}}\label{intro}

The outer automorphism groups of the ADE Lie algebras are $\mathbb{Z}_2$ for $J=A_{N-1},\,D_{N\gt4}$ and $E_6$. But, for $J=D_4$, the group of outer automorphisms is the nonabelian group $S_3$.

When compactifying the $(2,0)$ theory of type $J$ on $C$, we can twist the compactification by an element of $Hom(\pi_1(C),Aut(J))$. When $Aut(J)$ is abelian, the homomorphism factors through $H_1(C)$ and so the possible twists are classified by $Hom(H_1(C),\mathbb{Z}_2)=H^1(C,\mathbb{Z}_2)$.

Our tinkertoy program has been based on chopping $C$ up into simple pieces (3-punctured spheres and cylinders) and classifying the possible theories corresponding to the pieces. This works well, even in the $\mathbb{Z}_2$-twisted case, essentially because there is a Mayer-Vietoris principle for $H^1(\bullet,\mathbb{Z}_2)$: one can understand the twisted theories on $C$ by understanding the twists of the component pieces of $C$.

This is no longer true in nonabelian case. There is no Mayer-Vietoris principle for $Hom(\pi_1(C),S_3)$, and this makes our classification strategy ineffective.

A simple example of the problem will suffice. We can form a genus-2 surface by gluing together two once-punctured tori along a circle $S$. The twist $\gamma$ around this (homologically trivial, but homotopically nontrivial) cycle must be trivial in the abelian case but can be \emph{nontrivial} in the nonabelian case. (For $S_3$, you can prove that $\gamma$ has order-3.)

But now think about constructing those once-punctured tori by taking a 3-punctured sphere and gluing two of the punctures together.

You might think that, if you want to sew them together, the twists around the two punctures must be inverses of each other (ie $\gamma_1$ and $(\gamma_1)^{-1}$ ). And you would be \emph{correct}. You might also think that, on an \emph{unsewn} 3-punctured sphere, if two of the twists are $\gamma_1$ and $(\gamma_1)^{-1}$, then the 3rd twist must be trivial. Again, you would be \emph{correct}. However, as soon as you connect the two punctures, you introduce a new cycle and the corresponding twist ($\gamma_2$). Because $S_3$ is non-abelian, the twist around the remaining puncture can now be nontrivial (it is equal to the group commutator $(\gamma_2)^{-1} (\gamma_1)^{-1} \gamma_2 \gamma_1$).

In short, just because the product of the three twists on the 3-punctured sphere is trivial does \emph{not} mean that the twist around the puncture on a 1-punctured torus has to be trivial. Sewing together 3-punctured spheres does \emph{not} capture the twist information you need to construct higher-genus surfaces -- essentially because there's no Mayer-Vietoris for homotopy.

Because of this difficulty, we will not attempt to study the full $S_3$-twisted $D_4$ theory in this paper (see, however, \cite{Tachikawa:2010vg,Agarwal:2013uga} for some preliminary work in this direction). Instead, we will focus on abelian subgroups of twists. In \cite{Chacaltana:2013oka}, we studied the $\mathbb{Z}_2\subset S_3$ twists of the $D_4$ theory. Here we will study the $\mathbb{Z}_3\subset S_3$ twists.

As this subgroup of twists is abelian, they are classified by $H^1(C,\mathbb{Z}_3)$. Denoting the generator of $\mathbb{Z}_3$ by $\omega$ (we'll use a multiplicative notation for $\mathbb{Z}_3$), the twisted 3-punctured spheres will (up to permutations of the punctures or replacing $\omega\leftrightarrow\omega^2$) come in two types: $1-\omega-\omega^2$ and $\omega-\omega-\omega$. A twisted cylinder connects a puncture of type $\omega$ with a puncture of type $\omega^2$.

Recall that one of the complications of the $\mathbb{Z}_2$-twisted $D_N$ theories (for $N$ even) was that the $\mathbb{Z}_2$ outer automorphism acted nontrivially on the set of nilpotent orbits of the untwisted theory. In particular \cite{Chacaltana:2013oka} the two nilpotent orbits (which we denoted in red and blue) corresponding to a very-even partition are exchanged by the $\mathbb{Z}_2$ action. This led to an additional ramification of the moduli space of the gauge theory: dragging a very-even puncture around a twisted puncture changed it from red to blue (and vice versa).

The same is true for the $\mathbb{Z}_3$ twist, except that there is a triple of punctures (which we denote by red, blue and green) which are cyclically permuted by dragging them around a twisted puncture\footnote{The particular $\mathbb{Z}_2$ outer automorphism of $D_4$ that we studied in \cite{Chacaltana:2013oka} was the one that preserved the vector representation while exchanging the two spinor representations. We might call this $\mathbb{Z}_2\subset S_3$, ``$\mathbb{Z}_2^{\text{green}}$''. We could equally-well have considered a $\mathbb{Z}_2^{\text{red}}$ or $\mathbb{Z}_2^{\text{blue}}$ subgroup of $S_3$.} .

\subsection{{$k$-differentials}}\label{differentials}

Just as the $k$-differentials $\phi_4$ and $\tilde{\phi}_4$ provide an eigenbasis of 4-differentials for the action of $\mathbb{Z}_2^{\text{green}}$, there is a corresponding eigenbasis of 4-differentials for the action of the two non-trivial elements of $\mathbb{Z}_3$, $\omega$ and $\omega^2$.

We denote this basis of $k$-differentials by $\{\phi_2,\phi^{(\omega^2)}_4,\phi^{(\omega)}_4,\phi'_6\}$, where:

\begin{equation}
\begin{aligned}
\phi_4^{(\omega^2)} &\equiv \phi'_4 - 2\,\sqrt{3}i\tilde{\phi}_4\\
\phi_4^{(\omega)} &\equiv \phi'_4 + 2\,\sqrt{3}i\tilde{\phi}_4\\
\phi'_6 & \equiv \phi_6 - \tfrac{1}{6}\phi_2\phi'_4
\end{aligned}
\label{newkdifferentials}\end{equation}
and $\phi'_4 \equiv \phi_4 - \tfrac{1}{4}(\phi_2)^2$. Correspondingly, we label the pole structure as $\{p_2,p^{(\omega^2)}_4,p^{(\omega)}_4,p'_6\}$.

\section{{Tinkertoys}}\label{tinkertoys}

\subsection{{Regular punctures}}\label{regular_punctures}

The table of $\omega$-twisted punctures is the following.

{\footnotesize
\renewcommand{\arraystretch}{2.25}

\begin{longtable}{|c|c|c|c|c|c|}
\hline
\mbox{\shortstack{Nahm\\pole}}&Hitchin Pole&Pole structure&Constraints&
\mbox{\shortstack{Flavour\\Group}}&$(\delta n_{h},\delta n_{v})$\\
\hline
\endhead
$0$&$G_2$&$\{1,\frac{10}{3},\tfrac{11}{3},5\}$&$-$&$(G_2)_8$&$(112,107)$\\
\hline
$A_1$&$(G_2(a_1),S_3)$&$\{1,\frac{10}{3},\tfrac{8}{3},5\}$&-&$SU(2)_{14}$&$(102,100)$\\
\hline
$\widetilde{A}_1$&$(G_2(a_1),\mathbb{Z}_2)$&$\{1,\frac{10}{3},\tfrac{8}{3},5\}$&$c^{(6)}_5=8\,\sqrt{2}a^{(2)}_{5/3}\left(64\left(a^{(2)}_{5/3}\right)^2+c^{(4)}_{10/3}\right)$&$SU(2)_5$&$(93,92)$\\
\hline
$G_2(a_1)$&$G_2(a_1)$&$\{1,\frac{10}{3},\tfrac{8}{3},5\}$&$\begin{gathered}c^{(4)}_{10/3}=-48\left(\left(a^{(2)}_{5/3}\right)^2+3\left(a'^{(2)}_{5/3}\right)^2\right)\\ c^{(6)}_5=128\,\sqrt{2}a^{(2)}_{5/3}\left(\left(a^{(2)}_{5/3}\right)^2-9\left(a'^{(2)}_{5/3}\right)^2\right)\end{gathered}$&$-$&$(88,88)$\\
\hline
$G_2$&$0$&$\{1,\frac{7}{3},\frac{8}{3},4\}$&$\begin{gathered}c^{(4)}_{8/3}=-6\left(a^{(2)}_{4/3}\right)^2\\c^{(4)}_{7/3}=c^{(2)}_{1} a^{(2)}_{4/3} \\ c^{(6)}_{4}=-8\left(a^{(2)}_{4/3}\right)^3\\ c^{(6)}_{3}=\tfrac{1}{54}\left(c^{(2)}_1\right)^3+2a^{(2)}_{4/3}c^{(4)}_{5/3}\end{gathered}$&$-$&$(48,49)$\\
\hline
\end{longtable}
}

The $\omega^2$-twisted punctures are the same, but with $p^{(\omega^2)}_4\leftrightarrow p^{(\omega)}_4$.

\subsection{{Fixtures}}\label{fixtures}

In the following, we denote twisted punctures by their Bala-Carter labels and untwisted punctures by the corresponding partitions. Punctures in the $\omega$-twisted sector are in light-grey; punctures in the $\omega^2$-twisted sector are in dark-grey. We use $ \includegraphics[width=30pt]{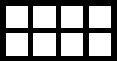}$ as a stand-in for the triple

\begin{displaymath}
 \includegraphics[width=137pt]{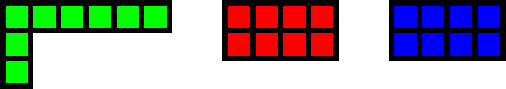}
\end{displaymath}
of untwisted punctures and $ \includegraphics[width=17pt]{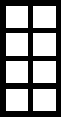}$ as a stand-in for the triple

\begin{displaymath}
 \includegraphics[width=137pt]{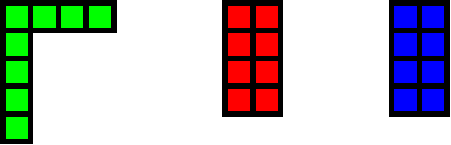}
\end{displaymath}
which are permuted by the $\mathbb{Z}_3$ action.

\subsection{{Free-field fixtures}}\label{freefield_fixtures}

\subsubsection{{$(1,\omega, \omega^2)$-Twisted Sector}}\label{twisted_sector}

\begin{longtable}{|c|c|c|c|}
\hline
\#&Fixture&Number of hypers&Representation\\
\hline 
\endhead
1&$\begin{matrix} \includegraphics[width=92pt]{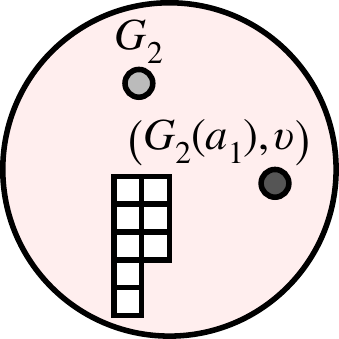}\end{matrix}$&0&empty\\
\hline
2&$\begin{matrix} \includegraphics[width=92pt]{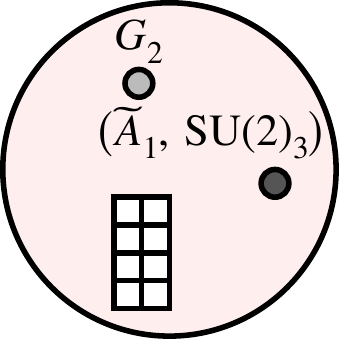}\end{matrix}$&$3$&$\tfrac{1}{2}(3,2)$\\
\hline
3&$\begin{matrix} \includegraphics[width=99pt]{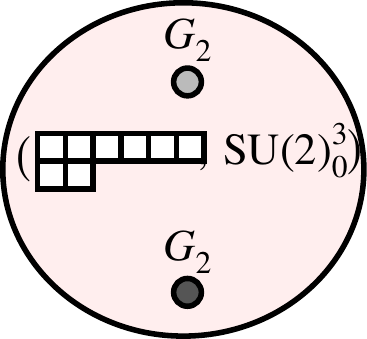}\end{matrix}$&0&empty\\
\hline
\end{longtable}

\subsubsection{{$(\omega,\omega,\omega)$-Twisted Sector}}\label{twisted_sector_2}

None.

\subsection{{Interacting Fixtures}}\label{interacting_fixtures}

\subsubsection{{$(1,\omega, \omega^2)$-Twisted Sector}}\label{twisted_sector_3}

\begin{longtable}{|c|c|c|c|c|}
\hline
$\#$&Fixture&$(d_2,d_3,d_4,d_6)$&$(n_h,n_v)$&$G_{\text{global}}$\\
\hline 
\endhead
1&$\begin{matrix} \includegraphics[width=92pt]{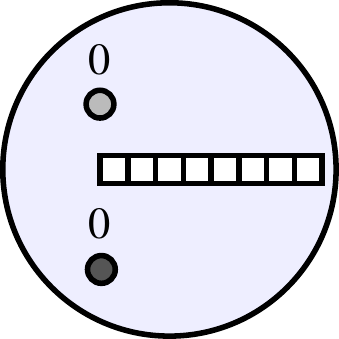}\end{matrix}$&$(0,0,6,4)$&$(112,86)$&$Spin(8)_{12}\times {(G_2)}^2_8$\\
\hline
2&$\begin{matrix} \includegraphics[width=92pt]{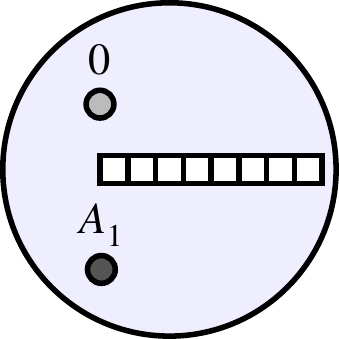}\end{matrix}$&$(0,0,5,4)$&$(102,79)$&$Spin(8)_{12}\times {(G_2)}_8\times SU(2)_{14}$\\
\hline
3&$\begin{matrix} \includegraphics[width=92pt]{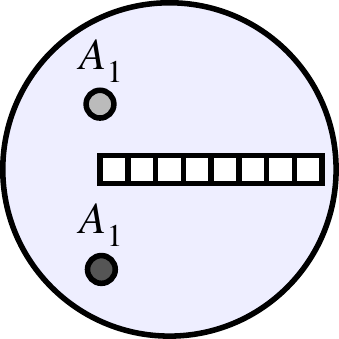}\end{matrix}$&$(0,0,4,4)$&$(92,72)$&$Spin(8)_{12}\times {SU(2)}^2_{14}$\\
\hline
4&$\begin{matrix} \includegraphics[width=92pt]{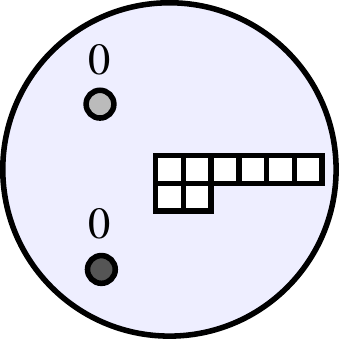}\end{matrix}$&$(0,0,6,3)$&$(96,75)$&${SU(2)}^3_8 \times {(G_2)}^2_8$\\
\hline
5&$\begin{matrix} \includegraphics[width=92pt]{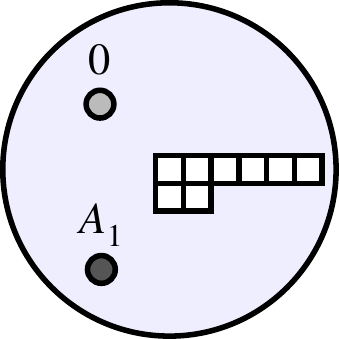}\end{matrix}$&$(0,0,5,3)$&$(86,68)$&${SU(2)}^3_8 \times {(G_2)}_8 \times SU(2)_{14}$\\
\hline
6&$\begin{matrix} \includegraphics[width=92pt]{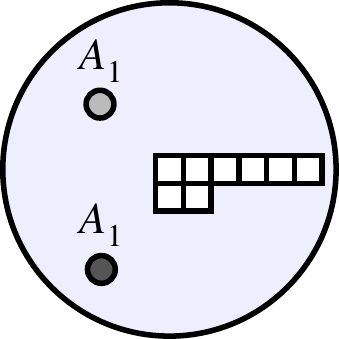}\end{matrix}$&$(0,0,4,3)$&$(76,61)$&${SU(2)}^3_8 \times {SU(2)}^2_{14}$\\
\hline
7&$\begin{matrix} \includegraphics[width=92pt]{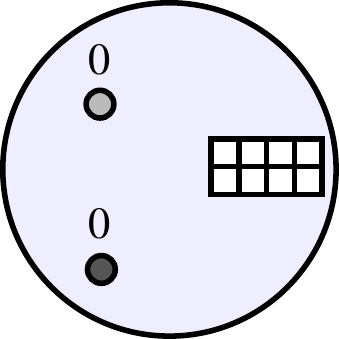}\end{matrix}$&$(0,0,5,3)$&$(88,68)$&$Sp(2)_8 \times {(G_2)}^2_8$\\
\hline
8&$\begin{matrix} \includegraphics[width=92pt]{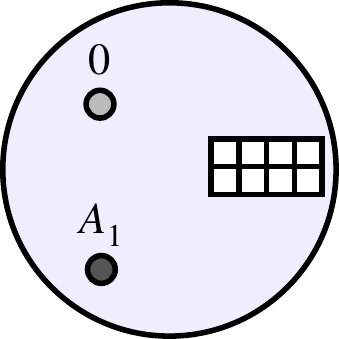}\end{matrix}$&$(0,0,4,3)$&$(78,61)$&$Sp(2)_8 \times {(G_2)}_8 \times SU(2)_{14}$\\
\hline
9&$\begin{matrix} \includegraphics[width=92pt]{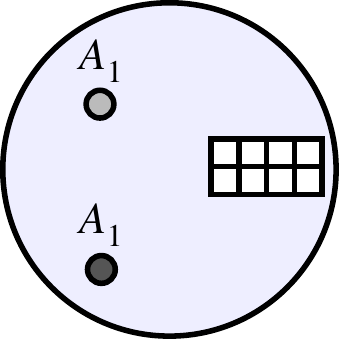}\end{matrix}$&$(0,0,3,3)$&$(68,54)$&$Sp(2)_8 \times {SU(2)}^2_{14}$\\
\hline
10&$\begin{matrix} \includegraphics[width=92pt]{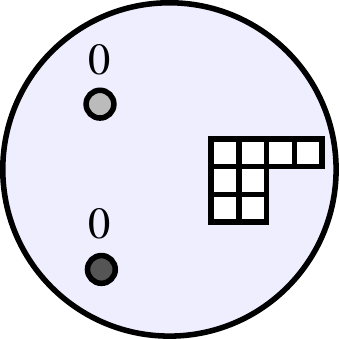}\end{matrix}$&$(0,1,4,2)$&$(72,55)$&${(G_2)}^2_8 \times {U(1)}^2$\\
\hline
11&$\begin{matrix} \includegraphics[width=92pt]{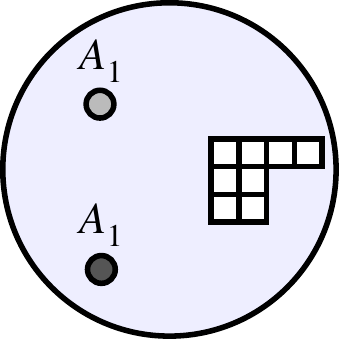}\end{matrix}$&$(0,1,3,2)$&$(62,48)$&${(G_2)}_8 \times SU(2)_{14} \times {U(1)}^2$\\
\hline
12&$\begin{matrix} \includegraphics[width=92pt]{f32120A1}\end{matrix}$&$(0,1,2,2)$&$(52,41)$&${SU(2)}^2_{14} \times {U(1)}^2$\\
\hline
12&$\begin{matrix} \includegraphics[width=92pt]{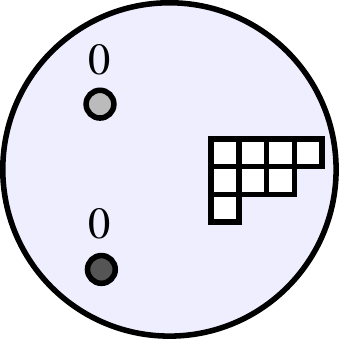}\end{matrix}$&$(0,0,4,3)$&$(79,61)$&$SU(2)_7\times {(G_2)}^2_8$\\
\hline
13&$\begin{matrix} \includegraphics[width=92pt]{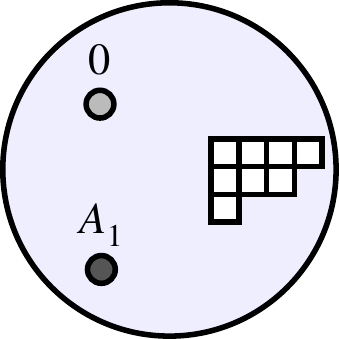}\end{matrix}$&$(0,0,3,3)$&$(69,54)$&$SU(2)_7 \times (G_2)_8 \times SU(2)_{14}$\\
\hline
14&$\begin{matrix} \includegraphics[width=92pt]{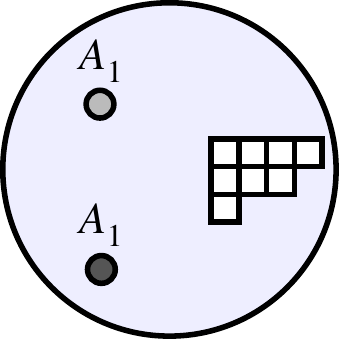}\end{matrix}$&$(0,0,2,3)$&$(59,47)$&$SU(2)_7 \times {SU(2)}^2_{14}$\\
\hline
15&$\begin{matrix} \includegraphics[width=92pt]{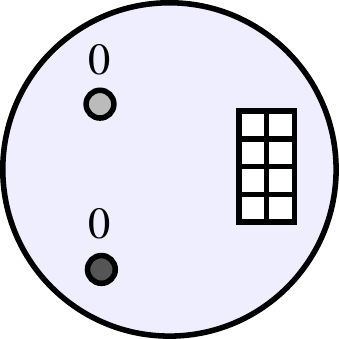}\end{matrix}$&$(0,0,3,1)$&$(48,32)$&$SU(2)_8 \times {(G_2)}^2_8$\\
\hline
16&$\begin{matrix} \includegraphics[width=92pt]{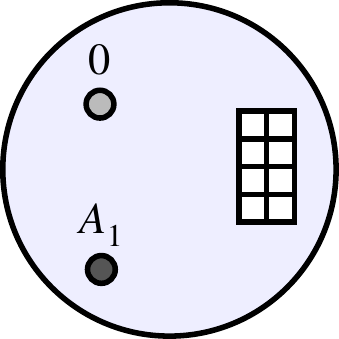}\end{matrix}$&$(0,0,2,1)$&$(38,25)$&$SU(2)_8 \times (G_2)_8 \times SU(2)_{14}$\\
\hline
17&$\begin{matrix} \includegraphics[width=92pt]{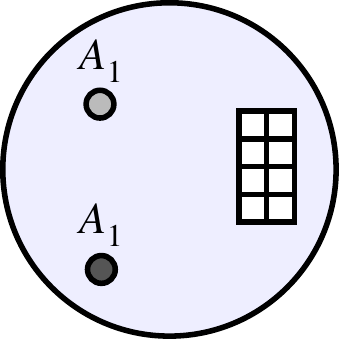}\end{matrix}$&$(0,0,1,1)$&$(28,18)$&$SU(2)_8 \times Sp(2)_{14}$\\
\hline\hypertarget{G2squared}
18&$\begin{matrix} \includegraphics[width=92pt]{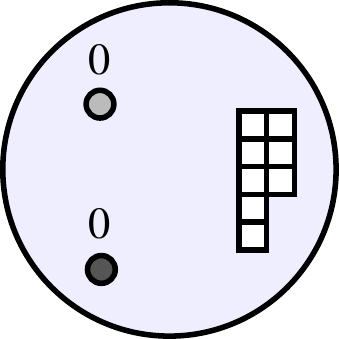}\end{matrix}$&$(0,0,2,1)$&$(40,25)$&${(G_2)}^2_8$\\
\hline
19&$\begin{matrix} \includegraphics[width=92pt]{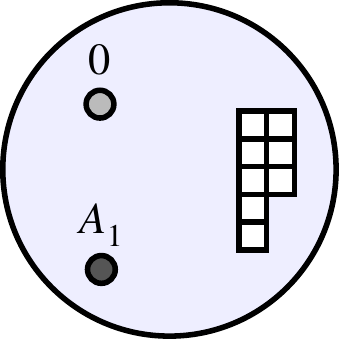}\end{matrix}$&$(0,0,1,1)$&$(30,18)$&$(G_2)_8 \times SU(2)_{14}$\\
\hline
20&$\begin{matrix} \includegraphics[width=92pt]{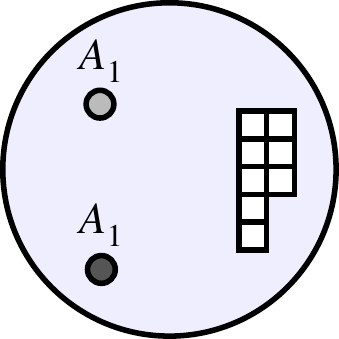}\end{matrix}$&$(0,0,0,1)$&$(20,11)$&$SU(4)_{14}$\\
\hline
\end{longtable}

\subsubsection{{$(\omega,\omega,\omega)$-Twisted Sector}}\label{twisted_sector_4}

\begin{longtable}{|c|c|c|c|c|}
\hline
$\#$&Fixture&$(d_2,d_3,d_4,d_6)$&$(n_h,n_v)$&$G_{\text{global}}$\\
\hline 
\endhead
1&$\begin{matrix} \includegraphics[width=92pt]{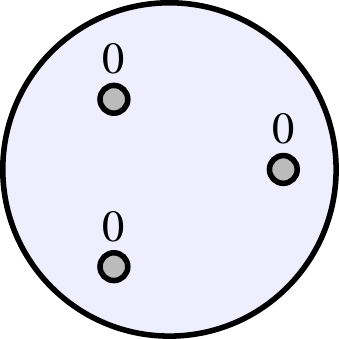}\end{matrix}$&$(0,0,7,4)$&$(112,93)$&${(G_2)}^3_8$\\
\hline
2&$\begin{matrix} \includegraphics[width=92pt]{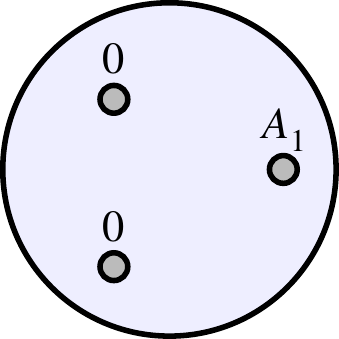}\end{matrix}$&$(0,0,6,4)$&$(102,86)$&${(G_2)}^2_8 \times SU(2)_{14}$\\
\hline
3&$\begin{matrix} \includegraphics[width=92pt]{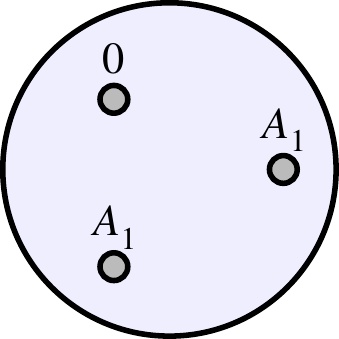}\end{matrix}$&$(0,0,5,4)$&$(92,79)$&${(G_2)}_8 \times {SU(2)}^2_{14}$\\
\hline
4&$\begin{matrix} \includegraphics[width=92pt]{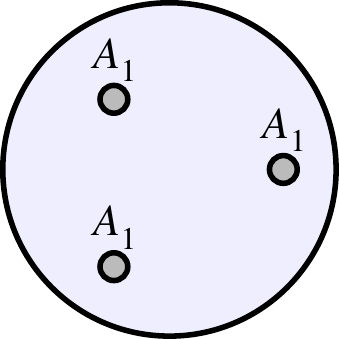}\end{matrix}$&$(0,0,4,4)$&$(82,72)$&${SU(2)}^3_{14}$\\
\hline
\end{longtable}

\subsection{{Gauge Theory Fixtures}}\label{gauge_theory_fixtures}

The punctures $G_2, G_2(a_1),$ and $\tilde{A}_1$ are atypical punctures, in the sense that they contribute $d_2\gt 1$ to the Coulomb branch dimension (they contribute $d_2=2,3,2$ respectively). As such, they ought to be ``resolved'' as the nonsingular OPE of two (or, in the case of $G_2(a_1)$, three) punctures. \footnote{For more details, see \cite{Chacaltana:2012ch}.} 

We can \emph{partially} resolve $G_2(a_1)$, while staying within the sector of commuting twists: $G_2(a_1)$ is the nonsingular OPE of $G_2$ with the simple puncture, $[5,3]$, from the untwisted sector.

However, resolving $G_2$ requires leaving the tractable subset of commuting twists. $G_2$ is the OPE of two simple punctures from non-commuting $\mathbb{Z}_2$ twisted sectors

\begin{displaymath}
G_2 \sim [6]_v \cdot [6]_s
\end{displaymath}
Similarly,

\begin{displaymath}
\tilde{A}_1 \sim [6]_v\cdot [4,1^2]_s
\end{displaymath}
This poses a conundrum. If we want to explore the full structure of the space of theories which include these punctures, we need to consider the full nonabelian $S_3$ group of twists. If we don't, we are stuck with including the unresolved gauge theory fixtures. With the exception of \S\ref{resolving_the_atypical_punctures}, we will restrict ourselves to commuting twists and hence will leave these atypical punctures unresolved.

\subsubsection{{$(1,\omega, \omega^2)$-Twisted Sector}}\label{twisted_sector_5}

There are 50 gauge theory fixtures, and one more with with an irregular puncture

\begin{longtable}{|c|c|c|c|c|c|}
\hline
\#&Fixture&$(d_2,d_3,d_4,d_6)$&$G$&Number of Hypers&Representation\\
\hline 
\endhead
1&$\begin{matrix}\includegraphics[width=84pt]{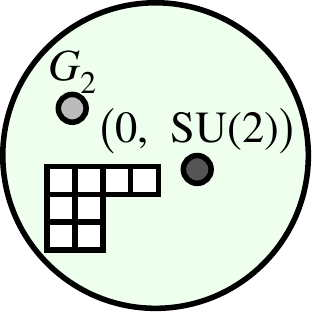}\end{matrix}$&$(1,0,0,0)$&$SU(2)$&$8$&$\begin{gathered}(3;2)\\+(1;2)\end{gathered}$\\
\hline
\end{longtable}

Of the 50 with three regular punctures, those with enhanced global symmetry are

{\footnotesize
\renewcommand{\arraystretch}{2.25}

\begin{longtable}{|c|c|c|c|c|c|c|}
\hline
$\#$&Fixture&$(d_2,d_3,d_4,d_6)$&$G$&$(n_h,n_v)$&\mbox{\shortstack{Rep of \\$G_\text{global} \times G$}}&$G_\text{global}$\\
\hline 
\endhead
1&$\begin{matrix} \includegraphics[width=84pt]{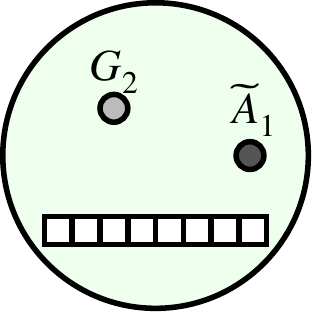}\end{matrix}$&$(2,0,1,0)$&$SU(2) \times Sp(2)$&$(29,13)$&$\begin{gathered}\frac{1}{2}(8_v,1,1;2,1)\\ +\frac{1}{2}(1,8_s,1;1,4)\\ +\frac{1}{2}(1,1,2;1,5)\end{gathered}$&$\begin{gathered}Spin(8)_4 \times\\ Spin(8)_{8} \times SU(2)_5\end{gathered}$\\
\hline
2&$\begin{matrix} \includegraphics[width=84pt]{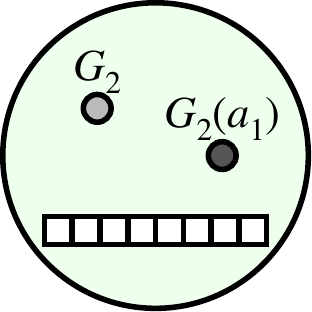}\end{matrix}$&$(3,0,0,0)$&${SU(2)}^3$&$(24,9)$&$\begin{gathered}\frac{1}{2}(8_v,1,1;2,1,1)\\+\frac{1}{2}(1,8_v,1;1,2,1)\\ +\frac{1}{2}(1,1,8_s;1,1,2)\end{gathered}$&${Spin(8)}^3_{4}$\\
\hline
3&$\begin{matrix}  \includegraphics[width=84pt]{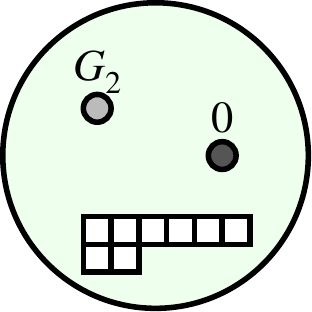}\end{matrix}$&$(1,0,2,0)$&$Sp(2)$&$(32,17)$&$\begin{gathered}\frac{1}{2}(1,2,2;4)\\ +[(E_7)_{8} SCFT]\end{gathered}$&$Spin(7)_8 \times {SU(2)}^3_8$\\
\hline
4&$\begin{matrix} \includegraphics[width=84pt]{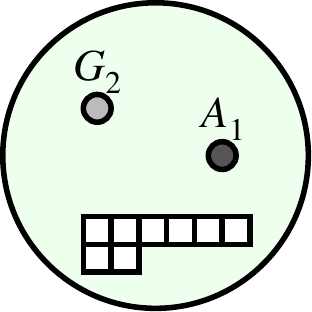}\end{matrix}$&$(1,0,1,0)$&$Sp(2)$&$(22,10)$&$\begin{gathered}\frac{1}{2}(1,1,2;1)\\+\frac{1}{2}(8_v,1,1;4)\\+\frac{1}{2}(1,2,1;5)\end{gathered}$&$\begin{gathered}Spin(8)_8 \times\\ SU(2)_5 \times SU(2)_1\end{gathered}$\\
\hline
5&$\begin{matrix} \includegraphics[width=84pt]{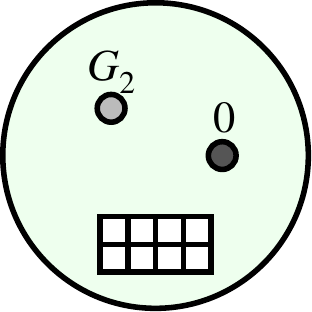}\end{matrix}$&$(1,0,1,0)$&$Sp(2)$&$(24,10)$&$\frac{1}{2}(12;4)$&$Spin(12)_{8}$\\
\hline
6&$\begin{matrix} \includegraphics[width=84pt]{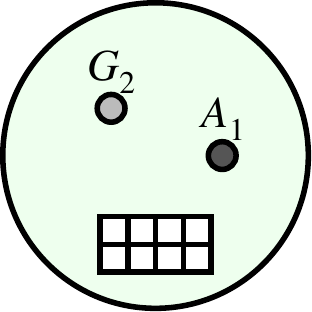}\end{matrix}$&$(1,0,0,0)$&$SU(2)$&$(14,3)$&$\frac{1}{2}(1,12;1)+\frac{1}{2}(8_v,1;2)$&$Spin(8)_4 \times Sp(6)_4$\\
\hline
7&$\begin{matrix} \includegraphics[width=84pt]{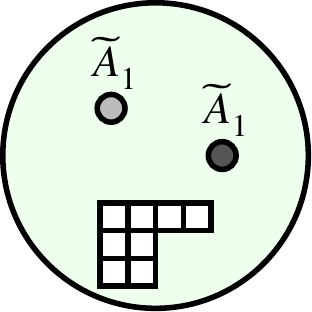}\end{matrix}$&$(2,1,2,0)$&$Sp(2) \times Sp(2)$&$(34,25)$&$\begin{gathered}\frac{1}{2}(2,1;5,1)+\frac{1}{2}(1,2;1,5)\\+(1,1;4,1)+(1,1;1,4)\\ +[(E_6)_6 \, \text{SCFT}]\end{gathered}$&${SU(2)}^2_5 \times {U(1)}^3$\\
\hline
8&$\begin{matrix} \includegraphics[width=84pt]{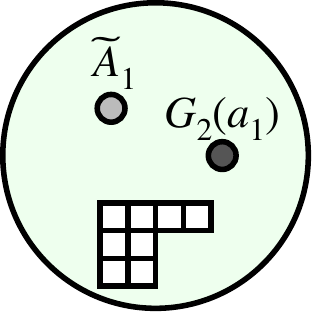}\end{matrix}$&$(3,1,1,0)$&$\begin{gathered}Sp(2) \times SU(2)\\ \times SU(3) \end{gathered}$&$(29,21)$&$\begin{gathered}\frac{1}{2}(2;5,1,1)+(1;4,1,1)\\+(1;1,2,1)+(1;4,1,3)\\+(1;1,2,3) \end{gathered}$&$SU(2)_5 \times {U(1)}^4$\\
\hline
9&$\begin{matrix} \includegraphics[width=84pt]{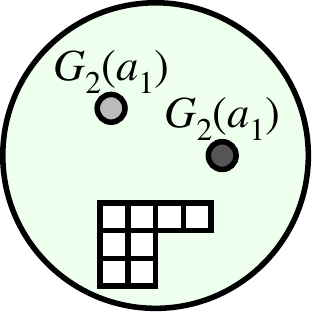}\end{matrix}$&$(4,1,0,0)$&$\begin{gathered}{SU(2)}^3\\\times SU(3)\end{gathered}$&$(24,17)$&$\begin{gathered}(2,1,1,1)+(1,2,1,1)\\+(1,1,2,1)+(2,1,1,3)\\+(1,2,1,3)+(1,1,2,3) \end{gathered}$&${U(1)}^6$\\
\hline
10&$\begin{matrix}\includegraphics[width=84pt]{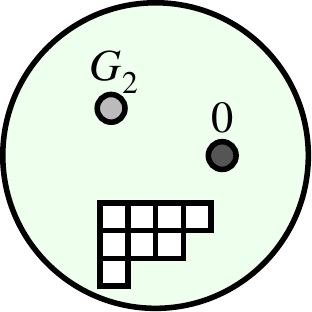}\end{matrix}$&$(1,0,0,0)$&$SU(2)$&$(15,3)$&$\begin{gathered}\frac{1}{2}(1,7,2;1)\\+\frac{1}{2}(8_v,1,1;2)\end{gathered}$&$\begin{gathered}Spin(8)_4 \times\\ Sp(7)_2 \times SU(2)_7\end{gathered}$\\
\hline
11&$\begin{matrix} \includegraphics[width=84pt]{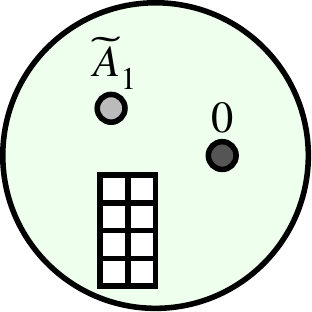}\end{matrix}$&$(1,0,2,0)$&$Sp(2)$&$(29,17)$&$\begin{gathered}\frac{1}{2}(1,1,2;5)\\+[(E_7)_8 \, \text{SCFT}]\end{gathered}$&$\begin{gathered}SU(2)_8 \times \\ Spin(7)_8 \times SU(2)_5\end{gathered}$\\
\hline
12&$\begin{matrix} \includegraphics[width=84pt]{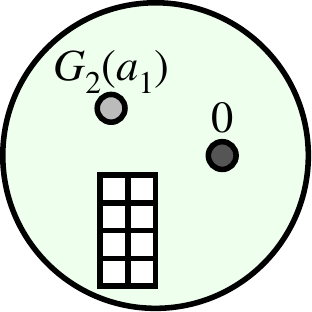}\end{matrix}$&$(2,0,1,0)$&$Sp(2) \times SU(2)$&$(24,13)$&$\begin{gathered}\frac{1}{2}(2,1;4,2)\\+\frac{1}{2}(1,8_v;4,1)\end{gathered}$&$SU(2)_8 \times Spin(8)_8$\\
\hline
13&$\begin{matrix} \includegraphics[width=84pt]{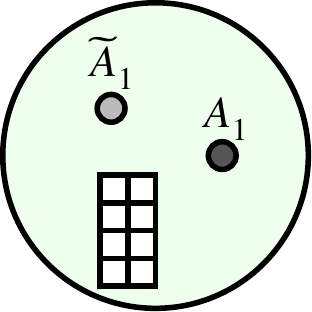}\end{matrix}$&$(1,0,1,0)$&$Sp(2)$&$(19,10)$&$\begin{gathered}\frac{1}{2}(1,1,1,2;1)\\+\frac{1}{2}(2,2,1,1;4)\\+\frac{1}{2}(1,1,4,1;5)\end{gathered}$&$\begin{gathered}{SU(2)}^2_8 \times\\ Sp(2)_5 \times SU(2)_1\end{gathered}$\\
\hline
14&$\begin{matrix} \includegraphics[width=84pt]{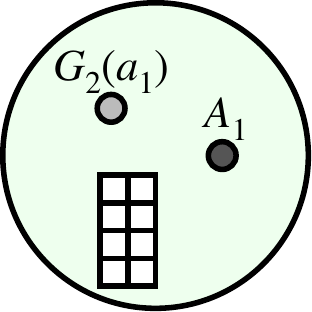}\end{matrix}$&$(2,0,0,0)$&$SU(2)\times SU(2)$&$(14,6)$&$\begin{gathered}\frac{1}{2}(1,1,1,1,1,2;1,1)\\+\frac{1}{2}(2,2,1,1,1,1;2,1)\\+\frac{1}{2}(1,1,2,2,1,1;1,2)\\+\frac{1}{2}(1,1,1,1,2,1;2,2) \end{gathered}$&${SU(2)}^5_4 \times SU(2)_2$\\
\hline
15&$\begin{matrix} \includegraphics[width=84pt]{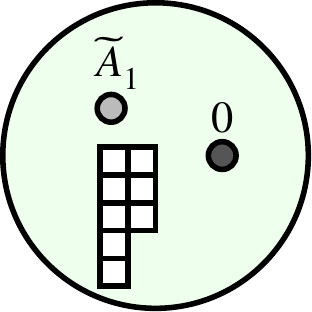}\end{matrix}$&$(1,0,1,0)$&$Sp(2)$&$(21,10)$&$\frac{1}{2}(8_v,1;4)+\frac{1}{2}(1,2;5)$&$Spin(8)_8 \times SU(2)_5$\\
\hline
16&$\begin{matrix} \includegraphics[width=84pt]{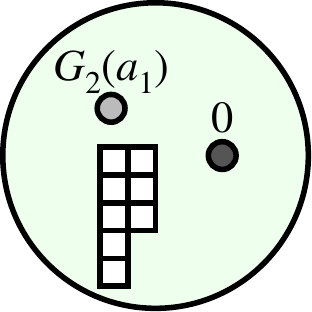}\end{matrix}$&$(2,0,0,0)$&$SU(2) \times SU(2)$&$(16,6)$&$\begin{gathered}\frac{1}{2}(8_v,1;2,1)\\+\frac{1}{2}(1,8_s;1,2)\end{gathered}$&${Spin(8)}^2_4$\\
\hline
17&$\begin{matrix} \includegraphics[width=84pt]{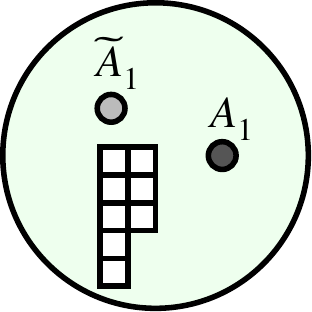}\end{matrix}$&$(1,0,0,0)$&$SU(2)$&$(11,3)$&$\frac{1}{2}(1,6;1)+\frac{1}{2}(8_v,1;2)$&$Spin(8)_4 \times Sp(3)_5$\\
\hline
\end{longtable}

}

\subsubsection{{$(\omega,\omega,\omega)$-Twisted Sector}}\label{twisted_sector_6}

There are 23 gauge theory fixtures. Those with enhanced global symmetry are

{\footnotesize
\renewcommand{\arraystretch}{2.25}

\begin{longtable}{|c|c|c|c|c|c|c|}
\hline
$\#$&Fixture&$(d_2,d_3,d_4,d_6)$&$G$&\mbox{\shortstack{Number\\of Hypers}}&
\mbox{\shortstack{Rep of\\$G_\text{global} \times G$}}&$G_\text{global}$\\
\hline 
\endhead
1&$\begin{matrix}\includegraphics[width=84pt]{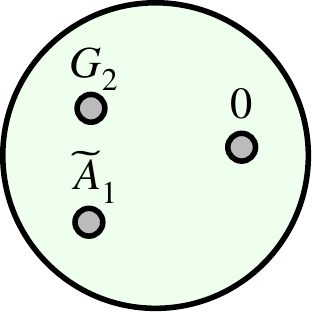}\end{matrix}$&$(2,0,2,0)$&$Sp(2) \times Sp(2)$&$(29,20)$&$\begin{gathered}\\\frac{1}{2}(1,2;5,1)\\+\frac{1}{2}(7,1;1,4)\\+\frac{1}{2}(1,1;5,4)\end{gathered}$&$Spin(7)_8 \times SU(2)_5$\\
\hline
2&$\begin{matrix}\includegraphics[width=84pt]{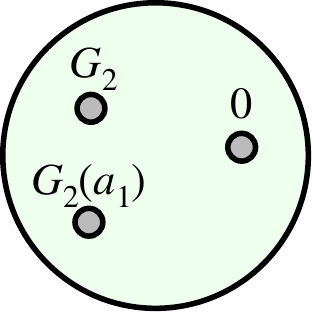}\end{matrix}$&$(3,0,1,0)$&$\begin{gathered}Sp(2) \times\\ SU(2) \times SU(2)\end{gathered}$&$(24,16)$&$\begin{gathered}\frac{1}{2}(1;4,2,2)\\+\frac{1}{2}(8_v;4,1,1)\end{gathered}$&$Spin(8)_8$\\
\hline
3&$\begin{matrix}\includegraphics[width=84pt]{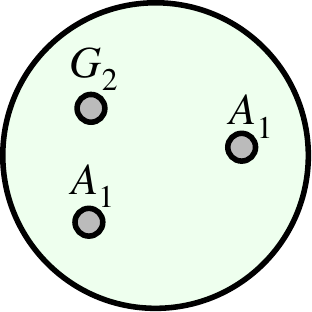}\end{matrix}$&$(1,0,1,1)$&$Sp(3)$&$(28,21)$&$\frac{1}{2}(4;14)$&${Sp(2)}_{14}$\\
\hline
4&$\begin{matrix}\includegraphics[width=84pt]{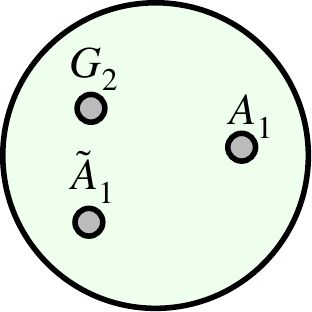}\end{matrix}$&$(2,0,1,0)$&$Sp(2) \times SU(2)$&$(19,13)$&$\begin{gathered}\frac{1}{2}(1,1,2;1,1)\\+\frac{1}{2}(1,3,1;1,2)\\+\frac{1}{2}(1,1,1;5,2)\\+\frac{1}{2}(4,1,1;5,1)\end{gathered}$&$\begin{gathered}Sp(2)_5 \times\\ SU(2)_8 \times SU(2)_1\end{gathered}$\\
\hline
5&$\begin{matrix}\includegraphics[width=84pt]{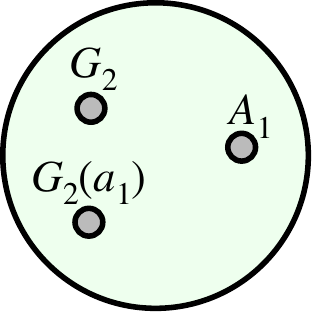}\end{matrix}$&$(3,0,0,0)$&${SU(2)}^3$&$(14,9)$&$\begin{gathered} \frac{1}{2}(2,1,1,1;2,2,1)\\+\frac{1}{2}(1,2,1,1;2,1,2)\\+\frac{1}{2}(1,1,2,1;1,2,2)\\+(1,1,1,2;1,1,1)\end{gathered}$&${SU(2)}^3_4 \times SU(2)_2$\\
\hline
\end{longtable}
}

\section{{Global symmetries and the superconformal index}}\label{global_symmetries_and_the_superconformal_index}

As we did for the $\mathbb{Z}_2$ twisted sector of the $D_4$ theory in \cite{Chacaltana:2013oka}, to determine the number of free hypermultiplets in each fixture and the global symmetry of each SCFT, we use the Hall-Littlewood limit of the superconformal index \cite{Gadde:2011ik,Kinney:2005ej,Gadde:2009kb,Gadde:2011uv,Lemos:2012ph}. When $C$ is a sphere, this limit of the index can be identified with the Hilbert series of the Coulomb branch of the $3d$ mirror of the $(2,0)$ theory compactified on $C \times S^1$ \cite{Benini:2010uu,Gaiotto:2012uq}. The 3d mirror of a $D_4$ fixture is obtained by assigning the $3d$ $\mathcal{N}=4$ SCFT $T_\rho[\mathfrak{g}^\vee]$ to each puncture of type $\rho$ (where for $\rho$ of type $1,\omega$ (or $\omega^2$), $\mathfrak{g}^\vee=\mathfrak{so}(8)$, $\mathfrak{g}_2$, respectively), and gauging the diagonal $G^\vee$ flavor symmetry. The expression for the Coulomb branch Hilbert series is then easily obtained following \cite{Cremonesi:2014kwa,Cremonesi:2014vla}. The result for a fixture in each twisted sector is

\subsection{{$(1,\omega,\omega^2)$ twisted sector}}\label{}

In this case, the superconformal index is given by

\begin{displaymath}
\mathcal{I}=\mathcal{A}(\tau)\sum_{(a_1,a_2)}\frac{\mathcal{K}(\Lambda(a,\tau))P_{G_2}^{(a_1,a_2)}(\Lambda(a,\tau))\mathcal{K}(\Lambda(b,\tau))P_{G_2}^{(a_1,a_2)}(\Lambda(b,\tau))\mathcal{K}(\Lambda(c,\tau))P_{SO(8)}^{(a_2,a_1,a_2,a_2)}(\Lambda(c,\tau))}{P_{SO(8)}^{(a_2,a_1,a_2,a_2)}(\Lambda([7,1],\tau))}
\end{displaymath}

\subsection{{$(\omega,\omega,\omega)$ twisted sector}}\label{_2}

In this case, the superconformal index is given by

\begin{displaymath}
\mathcal{I}=\mathcal{A}(\tau)\sum_{(a_1,a_2)}\frac{\mathcal{K}(\Lambda(a,\tau))P_{G_2}^{(a_1,a_2)}(\Lambda(a,\tau))\mathcal{K}(\Lambda(b,\tau))P_{G_2}^{(a_1,a_2)}(\Lambda(b,\tau))\mathcal{K}(\Lambda(c,\tau))P_{G_2}^{(a_1,a_2)}(\Lambda(c,\tau))}{P_{SO(8)}^{(a_2,a_1,a_2,a_2)}(\Lambda([7,1],\tau))}
\end{displaymath}
where $(a_1,a_2)$ are the Dynkin labels of a finite-dimensional irreducible representation of $G_2$ and $P^\lambda_G$ is a Hall-Littlewood polynomial of type $G$. \footnote{For a discussion of the Macdonald limit of the superconformal index of $D_4$ fixtures with $\mathbb{Z}_3$ twists, see \cite{Agarwal:2013uga}.}  Here, $\Lambda(a,\tau)$ denotes the fugacities associated to a puncture of type $\Lambda$, while $\Lambda([7,1],\tau)$ deonotes the fugacities associated to the trivial puncture.

An explicit expression for the Hall-Littlewood polynomials of type $SO(8)$ can be found in section 2.6 of \cite{Chacaltana:2013oka}. The Hall-Littlewood polynomials of type $G_2$ are defined by

\begin{equation}
P^\lambda_{G_2}(x_1,x_2,x_3;\tau)=W^{-1}_{\lambda}(\tau)\sum_{\sigma \in D_6}{x_1}^{\sigma(\lambda_1)}{x_2}^{\sigma(\lambda_2)}{x_3}^{\sigma(\lambda_3)}\prod_{\alpha \in R^+}\frac{1-\tau^2 x_1^{\sigma(-\alpha_1)}x_2^{\sigma(-\alpha_2)}x_3^{\sigma(-\alpha_3)}}{1- x_1^{\sigma(-\alpha_1)}x_2^{\sigma(-\alpha_2)}x_3^{\sigma(-\alpha_3)}}
\label{HLG2}\end{equation}
Letting $\{e_i\}$ denote the standard orthonormal basis of $\mathbb{R}^3$, $R^+=\{\alpha_1e_1+\alpha_2e_2+\alpha_3e_3\}$ is the set of positive roots \footnote{In this paper, we use the conventions of LieART \cite{Feger:2012bs}.} , $\lambda=\lambda_1 e_1+\lambda_2 e_2 + \lambda_3 e_3$ with

\begin{displaymath}
\begin{split}
\lambda_1&=-a_2, \\
\lambda_2&=-a_1-a_2, \\
\lambda_3&=a_1+2a_2, \\
\end{split}
\end{displaymath}
where $(a_1,a_2)$ are the Dynkin labels, and $x_i\equiv e^{e_i}$.

The Weyl group of $G_2$ is the dihedral group of order 12, which has presentation

\begin{displaymath}
D_6=\langle r,s|r^2=s^2=(rs)^6=1\rangle
\end{displaymath}
where $r,s$ acting on $\mathbb{R}^3$ can be represented as

\begin{displaymath}
r=\begin{pmatrix}0&1&0\\1&0&0\\0&0&1\end{pmatrix}, \, s=\begin{pmatrix}-1/3&2/3&2/3\\2/3&2/3&-1/3\\2/3&-1/3&2/3\end{pmatrix}
\end{displaymath}
The normalization factor $W_\lambda(\tau)$ is defined by

\begin{displaymath}
W_\lambda(\tau)=\sqrt{\sum_{\sigma \lambda = \lambda}\tau^{2\ell(\sigma)}}
\end{displaymath}
where $\ell(\sigma)$ is the length of the Weyl group element $\sigma$. For $(a_1,a_2)=(0,0)$, $W_{(0,0)}(\tau)=1+2\tau^2+2\tau^4+2\tau^6+2\tau^8+2\tau^{10}+\tau^{12}$, for $(a_1,0), (0,a_2), (a_1,a_1)$, $W_{(a_1,0)}(\tau)=W_{(0,a_2)}(\tau)=W_{(a_1,a_1)}(\tau)=1+\tau^2$, and for $(a_1,a_2)$, $W_{(a_1,a_2)}(\tau)=1$.

\section{{A new rank-1 SCFT}}\label{newSCFT}

We find a new rank-1 SCFT by compactifying the $D_4$ theory on

\begin{equation}
 \includegraphics[width=92pt]{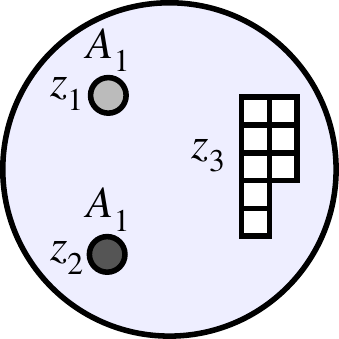}
\label{NewRank1SCFT}\end{equation}
The resulting theory has an ${SU(4)}_{14}$ global symmetry, and trace-anomaly coefficients $(n_h,n_v)=(20,11)$. The Coulomb branch is parametrized by a single complex scalar, with $\Delta(u)=6$.

It is a little surprising that the Seiberg-Witten curve, $\Sigma\subset T^*C$,

\begin{equation}
0 = \lambda^2 (\lambda^6 +\phi_6(z)),
\label{degen}\end{equation}
with

\begin{equation}
\phi_6(z)=\frac{u_6 z_{1 2}^4 z_{1 3} z_{2 3} {(dz)}^6}{{(z-z_1)}^5 {(z-z_2)}^5 {(z-z_3)}^2},
\label{phi6}\end{equation}
is reducible, even though the SCFT is not a product (of two SCFTs, or of an SCFT with free hypermultiplets). But this is not unprecedented. For instance, in the untwisted $D_4$ theory,

\begin{equation}
 \includegraphics[width=128pt]{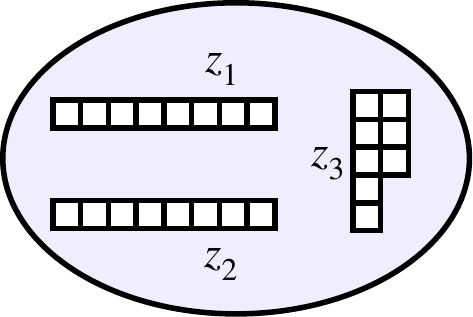}
\label{E8SCFT}\end{equation}
is a realization of the ${(E_8)}_{12}$ SCFT (which has $(n_h,n_v)=(40,11)$).

Its Seiberg-Witten curve is \emph{also} given by \eqref{degen},\eqref{phi6}. However, the set of mass deformations (which transform as the Cartan of the global symmetry of the fixture\footnote{Only the mass deformations, corresponding to the Cartan of the \emph{manifest} global symmetry of the fixture, are realizable in the Class-S realization. Thus we can realize only a 2-parameter family of mass deformations, even though $\text{rank}(SU(4))=3$.} ) is different.

The mass-deformed curve corresponding to \eqref{E8SCFT} is

\begin{equation}
0 = \lambda^8 +\lambda^6 \phi_2(z)+ \lambda^4 \phi_4(z) + \lambda^2\phi_6(z) + {\tilde{\phi}(z)}^2
\label{massdeformed}\end{equation}
where

\begin{displaymath}
\begin{aligned}
\phi_2(z)&=\frac{\left[m_2(z-z_1)z_{2 3}+m'_2(z-z_2)z_{1 3}\right]z_{1 2}\, {(dz)}^2}{{(z-z_1)}^2{(z-z_2)}^2{(z-z_3)}}\\
\phi_4(z)&=\frac{\left[\left(m_4(z-z_1)z_{2 3}-m'_4(z-z_2)z_{1 3}\right)(z-z_3)z_{1 2}+\tfrac{1}{4}{(m_2+m'_2)}^2 (z-z_1)(z-z_2)z_{1 3}z_{2 3}\right]z_{1 2}^2\, {(dz)}^4}{{(z-z_1)}^4{(z-z_2)}^4{(z-z_3)}^2}\\
\phi_6(z)&=\frac{\left[\left(m_6(z-z_1)z_{2 3}-m'_6(z-z_2)z_{1 3}\right)(z-z_3)z_{1 2}+u_6 (z-z_1)(z-z_2)z_{1 3}z_{2 3}\right]z_{1 2}^4\, {(dz)}^6}{{(z-z_1)}^6{(z-z_2)}^6{(z-z_3)}^2}\\
\tilde{\phi}(z)&=\frac{\left[\tilde{m}_4(z-z_1)z_{2 3}-\tilde{m}'_4(z-z_2)z_{1 3}\right]z_{1 2}^3\, {(dz)}^4}{{(z-z_1)}^4{(z-z_2)}^4{(z-z_3)}}
\end{aligned}
\end{displaymath}
It is useful to take the linear combinations defined in \eqref{newkdifferentials}:

\begin{displaymath}
\begin{aligned}
\phi'_4(z)&= \frac{\left[
\left( m_4+\tfrac{1}{4}   m_2^2\right)(z-z_1)z_{2 3}-
\left(m'_4+\tfrac{1}{4}{m'_2}^2\right)(z-z_2)z_{1 3}
\right]z_{1 2}^3\, {(dz)}^4}{{(z-z_1)}^4{(z-z_2)}^4{(z-z_3)}}
\end{aligned}
\end{displaymath}
which (like $\tilde{\phi}(z)$) has at worst a simple pole at the simple puncture (here, located at $z_3$), and $\phi'_6(z)=\phi_6- \tfrac{1}{6}\phi_2\phi'_4$. The latter, after mass-deforming, has a sextic pole at the untwisted full puncture but only a quintic pole at the twisted $A_1$ puncture.

In summary, mass-deforming the $A_1$ puncture in \eqref{NewRank1SCFT} allows $\phi_2(z)$ to have a double pole at the puncture, but \emph{does not} change the pole orders of $\phi_4^{(\omega)}$, $\phi_4^{(\omega^2)}$ or $\phi'_6$. Hence, the mass-deformed SW curve for the ${SU(4)}_{14}$ SCFT has

\begin{displaymath}
\phi_2(z)=\frac{\left[m_2(z-z_1)z_{2 3}+m'_2(z-z_2)z_{1 3}\right]z_{1 2}\, {(dz)}^2}{{(z-z_1)}^2{(z-z_2)}^2{(z-z_3)}}
\end{displaymath}
$\phi_6(z)$ as in \eqref{phi6}, and $\phi_4'(z)=\tilde{\phi}(z)=0$. That is, its mass deformations comprise a subspace of the space of mass deformations of \eqref{E8SCFT}.

On the basis of the Coulomb branch geometry alone, one would be hard-pressed to distinguish between the two theories. Nonetheless, they are obviously distinct. The Higgs branch of the ${(E_8)}_{12}$ SCFT is 29-dimensional, whereas the Higgs branch of this new one is 9-dimensional. \cite{Argyres:2015ffa,Argyres:2015gha} attempted to initiate a program of classifying rank-1 $\mathcal{N}=2$ SCFTs, based on their Coulomb branch geometry. As this example shows, such a classification may be incomplete.

The superconformal index of the ${SU(4)}_{14}$ SCFT is given by

\begin{equation}
\begin{split}
\mathcal{I}(a,b;\tau)&=\frac{(1-\tau^2)^2(1-\tau^4)(1-\tau^8)^2(1-\tau^{12})}{(1-\tau^2)(1-\tau^2 a^{\pm 2})(1-\tau^3 a^{\pm})(1-\tau^3 a^{\pm 3})(1-\tau^4)(1-\tau^2)}\\
&\times\frac{1}{
(1-\tau^2 b^{\pm 2})(1-\tau^3 b^{\pm})(1-\tau^3 b^{\pm 3})(1-\tau^4)(1-\tau^4)^3(1-\tau^6)(1-\tau^8)^2}\\
&\times \sum_{(a_1,a_2)}\frac{P_{G_2}^{(a_1,a_2)}(a\tau, 1, a^{-1}\tau;\tau)P_{G_2}^{(a_1,a_2)}(b\tau, 1, b^{-1}\tau;\tau)P^{(a_2,a_1,a_2,a_2)}_{SO(8)}(\tau^4,\tau^2,\tau^2,1;\tau)}{P^{(a_2,a_1,a_2,a_2)}_{SO(8)}(\tau^6,\tau^4,\tau^2,1;\tau)}
\end{split}
\label{SCInew}\end{equation}
The order $\tau^2$ expansion is given by

\begin{displaymath}
\mathcal{I}(a,b;\tau):=1+\chi^{\mathbf{15}}_{SU(4)}(a,b)\tau^2+\dots
\end{displaymath}
where

\begin{displaymath}
\chi^{\mathbf{15}}_{SU(4)}(a,b)=\chi^{\mathbf{3}}_{SU(2)}(a)+\chi^{\mathbf{3}}_{SU(2)}(b)+2\chi^{\mathbf{2}}_{SU(2)}(a)\chi^{\mathbf{2}}_{SU(2)}(b)+1
\end{displaymath}
Thus, the manifest global symmetry ${SU(2)}^2_{14}$ is enhanced to $SU(4)_{14}$.

We can study the Hall-Littlewood chiral ring \cite{Beem:2014rza} of this theory by taking the plethystic log of \eqref{SCInew}. This gives

\begin{equation}
PL[\mathcal{I}]=\chi^{\mathbf{15}}_{SU(4)}(a,b)\tau^2 + 2\chi^{\mathbf{20''}}_{SU(4)}(a,b)\tau^3+\chi^{\mathbf{50}}_{SU(4)}(a,b)\tau^4-2\chi^{\mathbf{20}}_{SU(4)}\tau^5-\dots
\label{PLnew}\end{equation}
From the above expression, the Hall-Littlewood chiral ring is generated by operators in the $\mathbf{15}$ of $SU(4)$ at order 2, two sets of operators in the $\mathbf{20''}$ at order 3, and operators in the $\mathbf{50}$ at order 4. These operators are subject to two relations at order 5, both transforming in the $\mathbf{20}$ of $SU(4)$. There are also higher-order relations, which can be extracted from the higher-order terms in \eqref{PLnew}.

\section{{The $T_{G_2}$ SCFT}}\label{_3}

In the untwisted compactifications of the $(2,0)$ theories of type $J$, a prominent role is played by the ``$T_J$ theory'', realized as the 3-punctured sphere with 3 full punctures. This theory has a ${(J)}^3_{2\kappa_J}$ global symmetry ($\kappa_J$ is the dual Coxeter number) and is, in a certain sense, the generic fixture. In particular, the theory associated to a genus-$g\geq 2$ Riemann surface, $C$, without punctures, is a ${(J)}^{3g-3}$ gauging of $(2g-2)$ copies of the $T_J$ theory. Different pants-decompositions of $C$ correspond to different S-duality frames of the gauge theory. The basic ``move,'' connecting the multitude of pants-decompositions, involves the three pants-decompositions of the 4-punctured sphere.

In the $\mathbb{Z}_2$-twisted theories, the same is true, but we need to also consider 4-punctured spheres with 2 or 4 full punctures from the $\mathbb{Z}_2$-twisted sector. Concomitantly, we encounter the fixture with 2 full punctures from the twisted sector and 1 full puncture from the untwisted sector:

\begin{equation}
 \includegraphics[width=108pt]{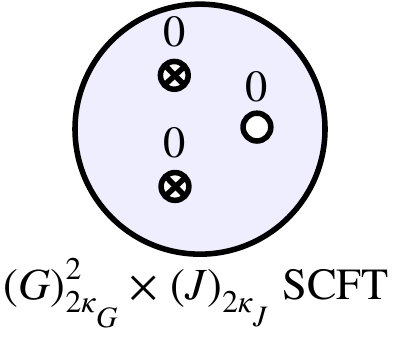}
\label{TGGU}\end{equation}
The basic duality (which allows one to pass between S-duality frames corresponding to different pants-decompositions of $C$) is

\begin{displaymath}
\begin{matrix} \includegraphics[width=202pt]{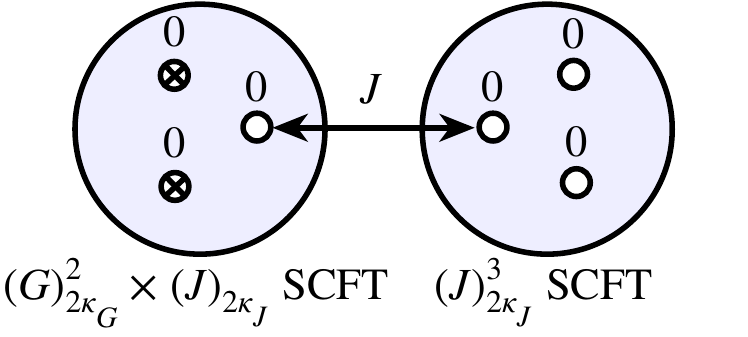}\end{matrix}\xLeftrightarrow{\qquad\qquad}\begin{matrix} \includegraphics[width=181pt]{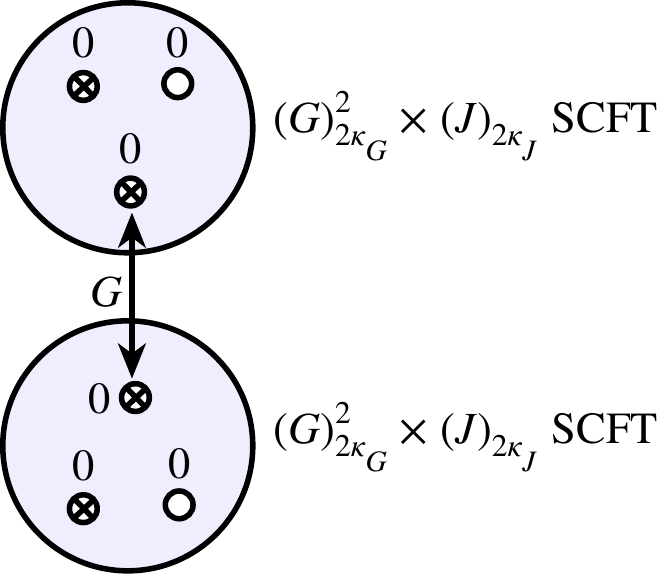}\end{matrix}
\end{displaymath}
(and a similar one, on the sphere with 4 twisted punctures, involving a $J$-gauging of two copies of \eqref{TGGU}).

In the $\mathbb{Z}_3$-twisted $D_4$ theory, we have, similarly,

\begin{displaymath}
\begin{matrix} \includegraphics[width=117pt]{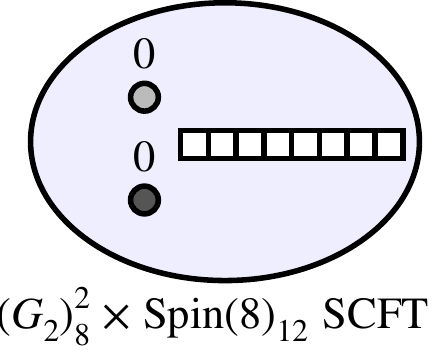}\end{matrix},\qquad
\begin{matrix} \includegraphics[width=75pt]{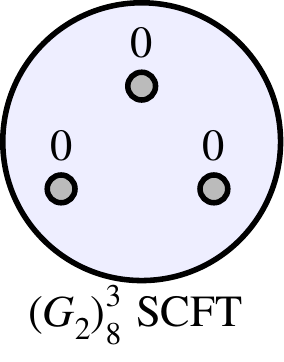}\end{matrix},\qquad \begin{matrix} \includegraphics[width=75pt]{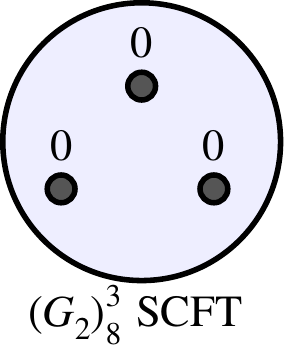}\end{matrix}
\end{displaymath}
The basic dualities are

\begin{displaymath}
\begin{matrix} \includegraphics[width=240pt]{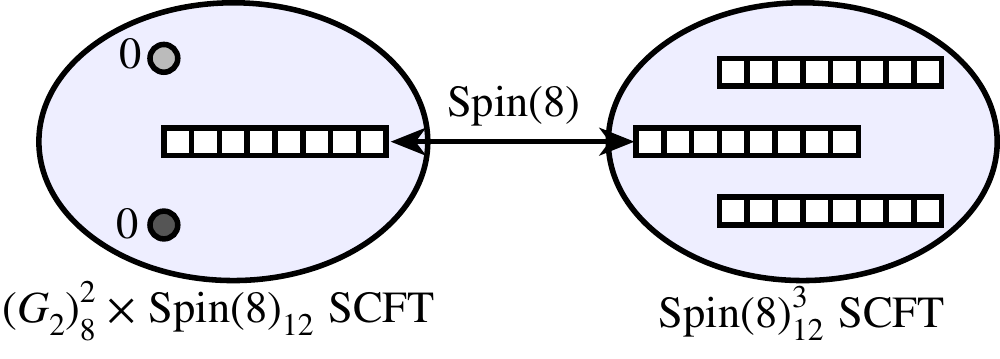}\end{matrix}\xLeftrightarrow{\qquad\qquad}\begin{matrix} \includegraphics[width=200pt]{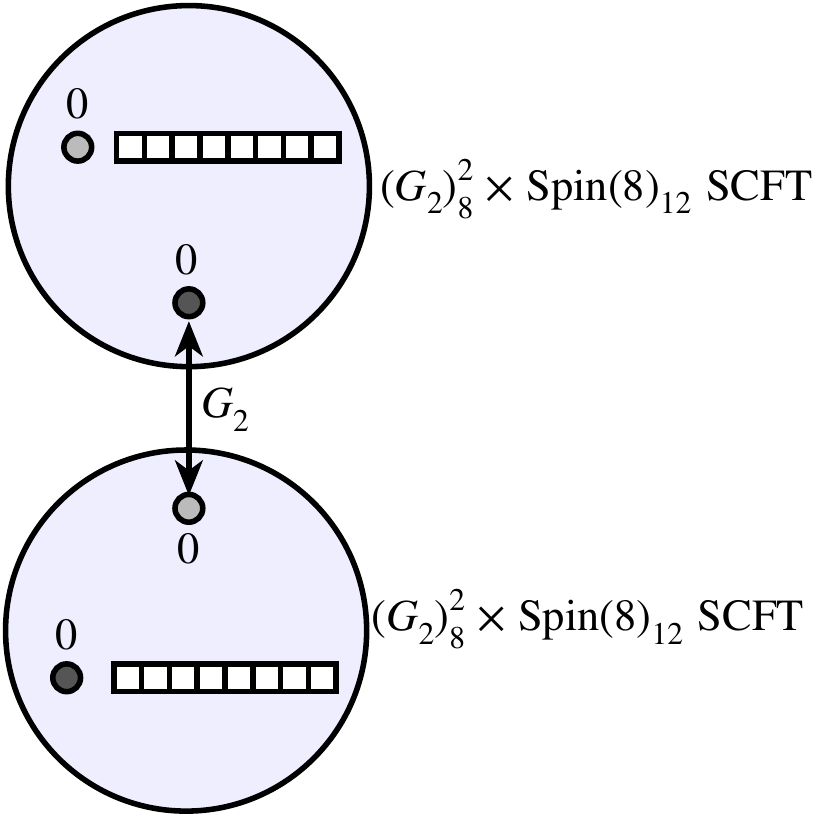}\end{matrix}
\end{displaymath}
and

\begin{displaymath}
\begin{matrix} \includegraphics[width=230pt]{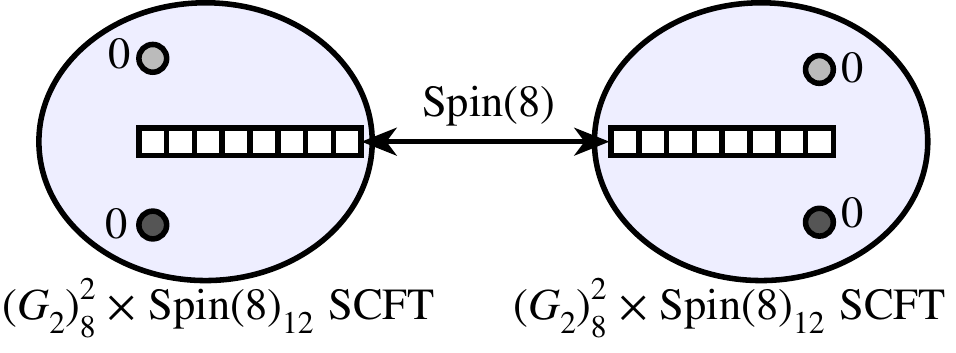}\end{matrix}\xLeftrightarrow{\qquad\qquad}
\begin{matrix} \includegraphics[width=124pt]{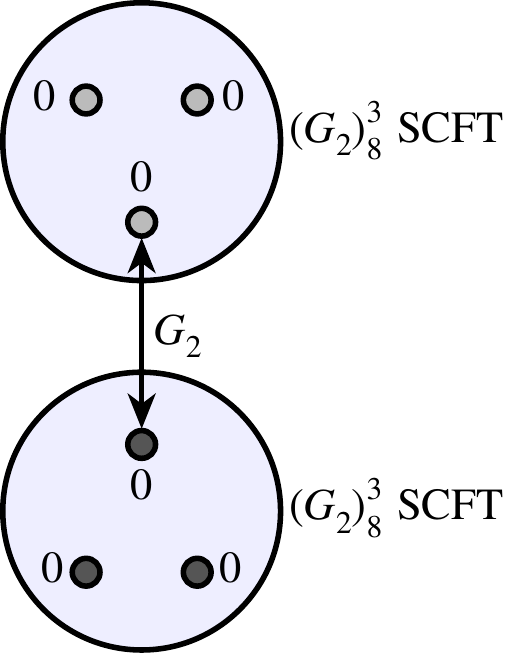}\end{matrix}
\end{displaymath}
and similarly for a 4 punctured sphere with one untwisted puncture and three twisted punctures from the \emph{same} twisted sector. The latter relate different $G_2$ gaugings of a product of the ${(G_2)}_8^3$ SCFT and the ${(G_2)}_8^2\times{Spin(8)}_{12}$ SCFT.

\section{{$Spin(8)$ gaugings of the ${(E_8)}_{12}$ SCFT}}\label{_gaugings_of_the__scft}

There are three inequivalent index-2 embeddings of $Spin(8)$ in $E_8$. They can be characterized by how the $248$ decomposes (up to outer automorphisms of $Spin(8)$). Either

\begin{subequations}
\begin{equation}
248 = 3(1) +5(28) +{35}_v + {35}_s +{35}_c
\label{E8decomp1}\end{equation}
or

\begin{equation}
248 = 1 +2(8_v)+3(28) +{35}_v + 2({56}_v)
\label{E8decomp2}\end{equation}
or

\begin{equation}
248 = 8_v+8_s+8_c+2(28) +{56}_v + {56}_s +{56}_c
\label{E8decomp3}\end{equation}
\end{subequations}
We can use one of these embeddings to gauge a $Spin(8)$ subgroup of the global symmetry group of the ${(E_8)}_{12}$ SCFT. The case of \eqref{E8decomp1} is realized in the untwisted $D_4$ theory as the once-punctured torus

\begin{displaymath}
 \includegraphics[width=167pt]{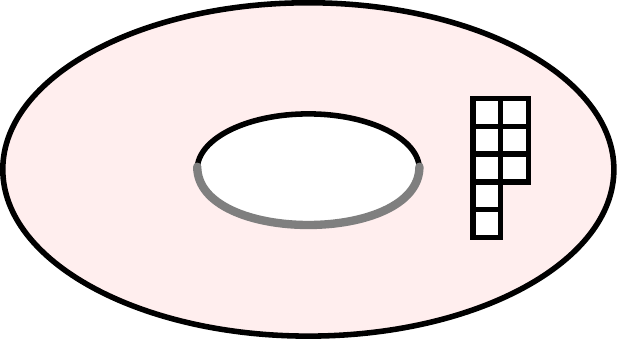}
\end{displaymath}
The gauge theory moduli space is the fundamental domain for $PSL(2,\mathbb{Z})$ in the UHP, and $\tau$ is the modular parameter of the torus.

As discussed in section 4.3 of \cite{Chacaltana:2013oka}, the case of \eqref{E8decomp2} is realized in the $\mathbb{Z}_2$-twisted $D_4$ theory. The S-dual theory is an $Sp(3)$ gauging of the ${Sp(6)}_8$ SCFT.

What about \eqref{E8decomp3}? That gauging can be realized in the $\mathbb{Z}_3$-twisted $D_4$ theory, considered here. We have $H^1(T^2-p,\mathbb{Z}_3)={(\mathbb{Z}_3)}^2$ and the action of $PSL(2,\mathbb{Z})$ consists of two orbits: the zero-orbit (corresponding to the untwisted theory) and the nonzero-orbit. The gauge theory moduli space is the fundamental domain for the index-8 subgroup, $\Gamma_1(3)\subset PSL(2,\mathbb{Z})$, which is the moduli space of pairs, $(C,\gamma)$, where $\gamma$ is a nonzero element of $H^1(T^2-p,\mathbb{Z}_3)$. The S-dual description\footnote{To be a bit more precise, the fundamental domain for $\Gamma_1(3)$ has \emph{3} cusp points, one of which corresponds to the point where the $Spin(8)$ gauging \eqref{E8decomp3} of the ${(E_8)}_{12}$ SCFT becomes weakly-coupled. The other two of correspond to points where a $G_2$ gauging of the ${(G_2)}_8^2$ SCFT becomes weakly-coupled.}  is a $G_2$ gauging of the ${(G_2)}_8^2$ SCFT (interacting fixture  \hyperlink{G2squared}{18}).

\section{{Resolving the atypical punctures}}\label{resolving_the_atypical_punctures}

As we have emphasized above, resolving the atypical punctures ($G_2,\, G_2(a_1)$ and $\tilde{A}_1$), from the $\mathbb{Z}_2$ twisted sector, requires stepping out of the world of commuting twists. Generically, this leads to a mess which resists any simple ``tinkertoy''-like classification. But, occasionally, one is lucky and can carry out a reasonable analysis, despite the non-commuting nature of the twists. In this section, we will give one example where such an analysis can be done.

Consider the 4-punctured sphere
\begin{equation}
 \includegraphics[width=107pt]{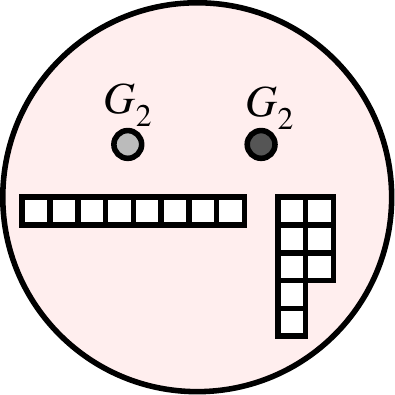}
\label{G2G25311111111}\end{equation}
Each $G_2$ puncture resolves into a pair of simple punctures from non-commuting $\mathbb{Z}_2$-twisted sectors. With a small loss of generality, we can therefore write this as the 6-punctured sphere

\begin{displaymath}
 \includegraphics[width=137pt]{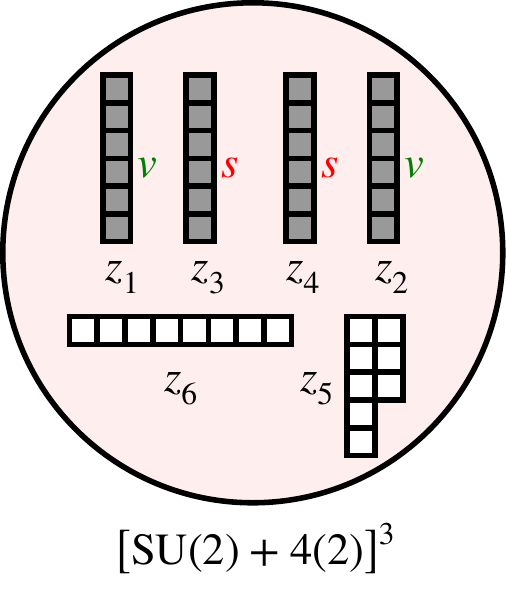}
\end{displaymath}
where we have attached a label( $\color{green}v$ or $\color{red}s$) to each twisted puncture, to indicate which $\mathbb{Z}_2$ subgroup of $S_3$ the twist belongs to.

The resulting theory is three decoupled copies of $SU(2)$ with 4 fundamentals. As such, we should be able to write down expressions for the $f(\tau_i)$ as functions on a branched cover of $\overline{M}_{0,6}$, branched over the compactification divisor. Unfortunately, this is too hard. It is easier, instead, to consider the $j$-invariant, $j(\tau_i)$, where

\begin{equation}
j(\tau)= \frac{4({f(\tau)}^2-f(\tau)+1)^3}{27 {f(\tau)}^2 {(1-f(\tau))}^2}
\end{equation}
We get a weakly-coupled gauge theory for $f(\tau)=0,1,\infty$; $j(\tau)$ has a double pole at all three of these points. Conversely, $j(\tau)$ has a triple zero at $f(\tau)= -\omega,-\omega^2$ (where $\omega^3=1$), and it is normalized so that $j(\tau)=1$ for $f(\tau) = -1,1/2,2$. So we seek three functions on a branched cover of $\overline{M}_{0,6}$ with double poles along appropriate components of the boundary and whose only zeroes are triple zeroes.

As a warmup, let us recall the example of section 5.1.2 of \cite{Chacaltana:2012ch}. There, we obtained ${(SU(N)+ 2N(\square))}^2$ gauge theory as the 5-punctured sphere

\begin{displaymath}
 \includegraphics[width=150pt]{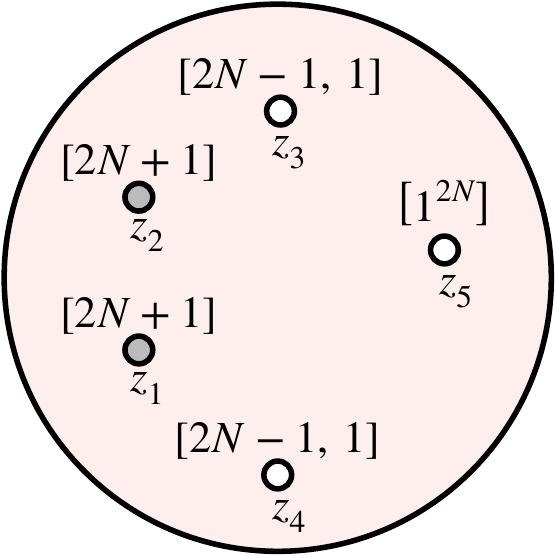}
\end{displaymath}
in the $\mathbb{Z}_2$-twisted $A_{2N-1}$ theory. Let

\begin{displaymath}
y_1^2 = \frac{z_{1 3}z_{2 5}}{z_{1 5}z_{2 3}},\qquad
y_2^2 = \frac{z_{1 4}z_{2 5}}{z_{1 5}z_{2 4}}
\end{displaymath}
which are single-valued on a 4-fold branched cover $X_5\to\overline{M}_{0,5}$. In \cite{Chacaltana:2012ch}, we showed that the gauge couplings

\begin{equation}
f(\tau_1) = \frac{y_1-1}{y_1+1}\frac{y_2-1}{y_2+1},\qquad
f(\tau_2) = \frac{y_1-1}{y_1+1}\frac{y_2+1}{y_2-1}
\end{equation}
Here, $f(\tau)=0,\infty$ is the weakly-coupled $SU(N)$ gauge theory; $f(\tau)=1$ is a weakly-coupled $SU(2)$ gauging of $1(2)+R_{2,N}$ SCFT. The corresponding $j$-invariants are

\begin{equation}
j(\tau_1) = \frac{{\bigl[{(1+y_1 y_2)}^2+3{(y_1+y_2)}^2\bigr]}^3}{27{\bigl(y_1^2-1\bigr)}^2{\bigl(y_2^2-1\bigr)}^2{\bigl(y_1+y_2\bigr)}^2}
\label{jtauM05}\end{equation}
(and $j(\tau_2)$ is given by $y_2\to -y_2$ in the above expression). For this family of gauge theories, where the physics is \emph{not} invariant under the $S_3$ which exchanges $f=0,1,\infty$, the $j$-invariant is not a particularly useful parameter (as inequivalent theories can have the same $j$-invariant). But, for $SU(2)+4(2)$, $S_3$ \emph{is} a symmetry and the $j$-invariant uniquely labels inequivalent theories.

Now, the $j(\tau_i)$ in \eqref{jtauM05} are still functions on the branched cover $X\to\overline{M}_{0,5}$. But $j(\tau_1)j(\tau_2)$ is (the pullback of) a meromorphic function on $\overline{M}_{0,5}$:

\begin{equation}
j(\tau_1)j(\tau_2) = \frac{{\bigl[{\bigl(x_1 x_2 +4(x_1+x_2)\bigr)}^2-48 x_1x_2\bigr]}^3}{729 x_1^4 x_2^4 {(x_1-x_2)}^2}
\end{equation}
where

\begin{displaymath}
x_1= y_1^2-1= \frac{z_{1 2}z_{3 5}}{z_{1 5}z_{2 3}},\qquad
x_2= y_2^2-1= \frac{z_{1 2}z_{4 5}}{z_{1 5}z_{2 4}}
\end{displaymath}
Returning to \eqref{G2G25311111111}, we can count the number of $SU(2)$`s which become weakly coupled at each irreducible component of the boundary of $M_{0,6}$. The boundary consist of 15 divisors, $D_{i j}$, where two punctures collide, and 10 divisors, $D_{i j k}$, where three punctures collide. Of course, having three of the six punctures collide is conformally equivalent to having the other three collide; so $D_{1 2 3}\simeq D_{4 5 6}$, etc. The $D_{i j}$ are del Pezzo surfaces, $dP_4$; the $D_{i j k}\simeq \mathbb{C P}^1\times \mathbb{C P}^1$.

We find

\begin{longtable}{|c|c|}
\hline
Divisor&Weakly-coupled gauge group\\
\hline
\endhead
$D_{1 2}$&$SU(2)$\\
\hline
$D_{3 4}$&$SU(2)$\\
\hline
$D_{1 3}$&$\emptyset$\\
\hline
$D_{2 3}$&$\emptyset$\\
\hline
$D_{1 4}$&$\emptyset$\\
\hline
$D_{2 4}$&$\emptyset$\\
\hline
$D_{1 5}$&$\emptyset$\\
\hline
$D_{2 5}$&$\emptyset$\\
\hline
$D_{3 5}$&$\emptyset$\\
\hline
$D_{4 5}$&$\emptyset$\\
\hline
$D_{1 6}$&${SU(2)}^3$\\
\hline
$D_{2 6}$&${SU(2)}^3$\\
\hline
$D_{3 6}$&${SU(2)}^3$\\
\hline
$D_{4 6}$&${SU(2)}^3$\\
\hline
$D_{5 6}$&${SU(2)}^3$\\
\hline
$D_{1 2 3}$&${SU(2)}^2$\\
\hline
$D_{1 2 4}$&${SU(2)}^2$\\
\hline
$D_{1 2 5}$&${SU(2)}^2$\\
\hline
$D_{1 2 6}$&${SU(2)}^2$\\
\hline
$D_{1 3 4}$&${SU(2)}^2$\\
\hline
$D_{2 3 4}$&${SU(2)}^2$\\
\hline
$D_{1 3 5}$&$\emptyset$\\
\hline
$D_{1 3 6}$&$\emptyset$\\
\hline
$D_{2 3 5}$&$\emptyset$\\
\hline
$D_{2 3 6}$&$\emptyset$\\
\hline
\end{longtable}

``Normally,'' precisely one simple factor in the gauge group becomes weakly-coupled at an irreducible component of the boundary. In this sense, only $D_{1 2}$ and $D_{3 4}$ are ``normal.'' All of the other components of the boundary are atypical.

Unlike the previous case, we don't have a detailed understanding of $X_6\xrightarrow{\pi}\overline{M}_{0,6}$, on which the gauge couplings are single-valued. The best we know how to do is to write down a formula for $j(\tau_1)j(\tau_2)j(\tau_3)$ which \emph{we will assume}\footnote{It is not totally obvious that this assumption is correct; nevertheless, the existence of a nice formula \eqref{jjjfinal} lends it some credence.} is (the pullback of) a single-valued meromorphic function on $\overline{M}_{0,6}$, whose only poles lie on those boundary divisors where some gauge couplings become weak, and whose zeroes are triple-zeroes. Moreover, it must be invariant under the action of the dihedral group, $D_4$, with generators

\begin{equation}
\begin{split}
\alpha\colon& (z_1,z_2,z_3,z_4)\mapsto (z_1,z_2,z_4,z_3)\\
\beta\colon& (z_1,z_2,z_3,z_4)\mapsto (z_2,z_1,z_3,z_4)\\
\gamma\colon& (z_1,z_2,z_3,z_4)\mapsto (z_3,z_4,z_1,z_2)\\
\end{split}
\label{dihedral}\end{equation}
Introducing the cross ratios

\begin{displaymath}
x_1=\frac{z_{1 2} z_{3 6}}{z_{1 6} z_{3 2}},\qquad
x_2=\frac{z_{1 2} z_{4 6}}{z_{1 6} z_{4 2}},\qquad
x_5=\frac{z_{1 2} z_{5 6}}{z_{1 5} z_{2 6}}
\end{displaymath}
the generators act as

\begin{displaymath}
\begin{split}
\alpha\colon&\quad x_1\leftrightarrow x_2,\qquad x_5\to x_5\\
\beta\colon&\quad x_1\to \frac{x_1}{x_1-1},\qquad x_2\to \frac{x_2}{x_2-1},\qquad x_5\to\frac{x_5}{x_5-1}\\
\gamma\colon&\quad x_1\to \frac{x_1-x_2}{x_1(1-x_2)},\qquad
x_2\to 1-\frac{x_2}{x_1},\qquad x_5 \to -\frac{(x_1-x_2)x_5}{x_2(x_1+x_5-x_1 x_5)}\\
\end{split}
\end{displaymath}
These conditions more-or-less pin down the desired expression. Normalizing $j(\tau_1)j(\tau_2)j(\tau_3)=1$ at the point $D_{1 3}\cap D_{2 4}\cap D_{1 3 6}$ (and points related to it by the action of the dihedral group generated by \eqref{dihedral}), we obtain

\begin{equation}\label{jjjfinal}
\begin{split}
j(\tau_1)j(\tau_2)j(\tau_3)=
\frac{1}{531441\, x_1^6 x_2^6 (x_1-x_2)^{12} x_5^{12}}\cdot
{\left[16(x_1-x_2)^2+x_1^2 x_2^2 +8 x_1x_2(2-x_1-x_2)\right]}^3&\\
\cdot{\left[(x_1-x_2)^2+16x_1^2 x_2^2 +8 x_1x_2(2-x_1-x_2)\right]}^3&\\
\cdot\Bigl[x_5^4(x_1-x_2)^2 + s\, x_1x_2(1-x_5)(x_5-x_1(x_5-1))(x_5-x_2(x_5-1))\Bigr]^3&
\end{split}
\end{equation}
with one unknown (nonzero) constant, $s$. This has

\begin{itemize}%
\item triple zeroes along divisors in the \emph{interior} of $M_{0,6}$,
\item a 12th-order pole along $D_{1 2}$, $D_{3 4}$, $D_{5 6}$, $D_{1 2 3}$, $D_{1 2 4}$, $D_{1 3 4}$, $D_{2 3 4}$, $D_{1 2 6}$, $D_{3 4 6}$,
\item a 6th-order pole along $D_{1 6}$, $D_{2 6}$, $D_{3 6}$, $D_{4 6}$
\item and no other poles.

\end{itemize}
The orders of these poles gives us \emph{some} information about the ramification of $X_6\xrightarrow{\pi}\overline{M}_{0,6}$.

For instance, along $D_{1 2}$ (and $D_{3 4}$), we expected a double pole (since one $SU(2)$ becomes weak), but obtained a 12th-order pole. So the ramification index must be 6. Along $D_{1 6}$ (and $D_{2 6}$, $D_{3 6}$ and $D_{4 6}$), we expected a 6th-order pole (since three $SU(2)$s become weak) and --- since that's what we obtained --- the covering must be unramified there. Along $D_{5 6}$ the ramification index appears to be 2 and, along $D_{1 2 3}$, $D_{1 2 4}$, $D_{1 3 4}$, $D_{2 3 4}$, $D_{1 2 6}$ and $D_{3 4 6}\,(=D_{1 2 5})$, the ramification index is 3.

Determining the constant $s$ requires some knowledge of locations the zeroes of the $j(\tau_i)$, which is beyond our current abilities.

\section*{Acknowledgements}\label{Acknowledgements}
\addcontentsline{toc}{section}{Acknowledgements}
We would like to thank Prarit Agarwal, Chris Beem, Michele del Zotto, Noppadol Mekareeya, Andy Neitzke, Yuji Tachikawa, and Fei Yan for helpful discussions. The work of J.D.~was supported in part by the National Science Foundation under Grant No. PHY-1316033. The work of A.T.~was supported in part by the National Research Foundation of Korea grants 2005-0093843, 2010-220-C00003 and 2012K2A1A9055280. The work of O.C.~was supported in part by the INCT-Matem\'atica and the ICTP-SAIFR in Brazil through a Capes postdoctoral fellowship. J.D.~would like to thank the Simons Center for Geometry and Physics for hospitality during the 2015 Simons Summer Workshop in Mathematics and Physics, where the bulk of this work was done.

\begin{appendices}

\section{}
\subsection{{Embeddings of $SU(2)$ in $G_2$}}\label{appendix_embeddings_of__in_}

{\renewcommand{\arraystretch}{1.5}

\begin{longtable}{|c|c|c|l|l|}
\hline
Bala-Carter&$\mathfrak{f}$&Embedding indices&$7$&$14$\\
\hline 
\endhead
$A_1$&$\mathfrak{su}(2)$&$(1,3)$&$(1;3)+(2;2)$&$(1;3)+(2;4)+(3;1)$\\
\hline 
$\widetilde{A}_1$&$\mathfrak{su}(2)$&$(3,1)$&$(2;2)+(3;1)$&$(1;3)+(3;1)+(4;2)$\\
\hline 
$G_2(a_1)$&$-$&$4$&$(1)+2(3)$&$3(3)+(5)$\\
\hline 
$G_2$&$-$&$28$&$7$&$(3)+(11)$\\
\hline
\end{longtable}
}

\subsection{{Projection matrices for $SO(8)$}}\label{projection_matrices_for_}

{\renewcommand{\arraystretch}{1.25}

\begin{longtable}{|c|c|c|}
\hline
Partition&$\mathfrak{f}$&Projection Matrix\\
\hline
\endhead
$[2^2,1^4]$&${\mathfrak{su}(2)}^3$&$\begin{pmatrix}1&2&1&1\\1&0&0&0\\0&0&1&0\\0&0&0&1\end{pmatrix}$\\
\hline
$[3,1^5]$&$\mathfrak{sp}(2)$&$\begin{pmatrix}2&2&1&1\\0&0&1&1\\0&1&0&0\end{pmatrix}$\\
\hline
$[2^4]_{r,b}$&$\mathfrak{sp}(2)$&$\begin{pmatrix}1&2&1&2\\1&0&1&0\\0&1&0&0\end{pmatrix}$\\
\hline
$[3^2,1^2]$&${\mathfrak{u}(1)}^2$&$\begin{pmatrix}2&4&2&2\\2&0&1&1\\0&0&1&-1\end{pmatrix}$\\
\hline
$[3,2^2,1]$&$\mathfrak{su}(2)$&$\begin{pmatrix}2&3&2&2\\0&1&0&0\end{pmatrix}$\\
\hline
$[5,1^3]$&$\mathfrak{su}(2)$&$\begin{pmatrix}4&6&3&3\\0&0&1&1\end{pmatrix}$\\
\hline
$[4^2]$&$\mathfrak{su}(2)$&$\begin{pmatrix}3&6&3&4\\1&0&1&0\end{pmatrix}$\\
\hline
$[5,3]$&$-$&$\begin{pmatrix}4&6&4&4\end{pmatrix}$\\
\hline
$[7,1]$&$-$&$\begin{pmatrix}6&10&6&6\end{pmatrix}$\\
\hline
\end{longtable}
}

\subsection{{Projection matrices for $G_2$}}\label{projection_matrices_for__2}

{\renewcommand{\arraystretch}{1.5}

\begin{longtable}{|c|c|c|}
\hline
Bala-Carter&$\mathfrak{f}$&Projection Matrix\\
\hline 
\endhead
$A_1$&$\mathfrak{su}(2)$&$\begin{pmatrix}1&2\\1&0\end{pmatrix}$\\
\hline
$\widetilde{A}_1$&$\mathfrak{su}(2)$&$\begin{pmatrix}2&3\\0&1\end{pmatrix}$\\
\hline
$G_2(a_1)$&$-$&$\begin{pmatrix}2&4\end{pmatrix}$\\
\hline
$G_2$&$-$&$\begin{pmatrix}6&10\end{pmatrix}$\\
\hline
\end{longtable}
}

\end{appendices}

\bibliographystyle{utphys}
\bibliography{ref}

\end{document}